\RequirePackage{fix-cm}
\documentclass[smallextended]{svjour3}       % onecolumn (second format)
\smartqed  % flush right qed marks, e.g. at end of proof
\usepackage{graphicx}

\usepackage{float}
\usepackage{epsf,exscale,times}
\usepackage{amsmath,amsfonts,amssymb,bm,graphicx}
\usepackage[usenames]{color}
\usepackage{listings}
\usepackage{latexsym}
\usepackage{color}
\definecolor{gray97}{gray}{.97}
\definecolor{gray75}{gray}{.75}
\definecolor{gray45}{gray}{.45}
\definecolor{gray35}{gray}{.35}

\usepackage{listings}
\lstnewenvironment{listing}[1][]
   {\lstset{#1}\pagebreak[0]}{\pagebreak[0]}

\lstdefinestyle{consola}
   {basicstyle=\scriptsize\bf\ttfamily,
    backgroundcolor=\color{gray75},
   }

\lstdefinestyle{C}
   {language=C,
   }

\newfloat{program}{tbphH}{lop}[section]
\floatname{program}{Program}

\journalname{Journal of Global Optimization }
\begin{document}

\title{An efficient implementation of parallel simulated annealing algorithm in GPUs\thanks{This work is partially supported by I-Math Consolider Project (Reference: COMP-C6-0393), by MICINN (MTM2010-21135-C02-01) and by Xunta de Galicia (Grant CN2011/004 cofunded with FEDER funds). The authors also acknowledge some ideas suggested by J.L. Fern\'andez (Autonomous University of Madrid).}
}
%\subtitle{Do you have a subtitle?\\ If so, write it here}

\titlerunning{Efficient SA implementation for GPUs}        % if too long for running head

\author{A.M. Ferreiro \and J.A. Garc\'ia \and J.G. L\'opez-Salas \and C. V\'azquez}

%\authorrunning{Short form of author list} % if too long for running head

\institute{A.M. Ferreiro, J.A. Garc\'ia, J.G. L\'opez-Salas, C. V\'azquez (Corresponding author) \at
              Department of Mathematics, Faculty of Informatics, Campus Elvi\~na s/n, 15071-A Coru\~na (Spain)\\
              Tel.: +34-981167000\\
              Fax: +34-981167160 \\
              \email{\{aferreiro, jagrodriguez, jose.lsalas, carlosv\}@udc.es} }

\date{Received: date / Accepted: date}
% The correct dates will be entered by the editor

\maketitle

\begin{abstract}
In this work we propose a highly optimized version of a simulated annealing (SA) algorithm adapted to the more recently developed Graphic Processor Units (GPUs). The programming has been carried out with CUDA toolkit, specially designed for Nvidia GPUs. For this purpose, efficient versions of SA have been first analyzed and adapted to GPUs. Thus, an appropriate sequential SA algorithm has been developed as starting point. Next, a straightforward asynchronous parallel version has been implemented and then a specific and more efficient synchronous version has been developed. A wide appropriate benchmark to illustrate the performance properties of the implementation has been considered. Among all tests, a classical sample problem provided by the minimization of the normalized Schwefel function has been selected to compare the behavior of the sequential, asynchronous and synchronous versions, the last one being more advantageous in terms of balance between convergence, accuracy and computational cost. Also the implementation of a hybrid method combining SA with a local minimizer method has been developed. Note that the generic feature of the SA algorithm allows its application in a wide set of real problems arising in a large variety of fields, such as biology, physics, engineering, finance and industrial processes.
\keywords{Global optimization \and simulated annealing \and parallel computing \and GPUs \and CUDA.}
% \PACS{PACS code1 \and PACS code2 \and more}
% \subclass{MSC code1 \and MSC code2 \and more}
\end{abstract}

\section{Introduction}
\label{intro}
In this work we consider the box-constrained global minimization problem:
\begin{equation}
\mathop{min}\limits_{\pmb x\in I} \, f (\pmb x),
\label{eq:min_pb}
\end{equation}
where $f$ is the cost function, $\pmb x=(x_1,\ldots, x_n)\in I\subset \mathbb{R}^n$, the search space being $I=I_1\times \ldots \times I_n$, where $I_k=[l_k,u_k]$, with $l_k, \, u_k\in \mathbb{R}$ for $k=1,...,n$.

This kind of problems arises in many fields of application, such as physics, finance, industry, biology, etc. Usually
the dimension of the optimization problem is very large, and the evaluation of the function involves
a high computational cost.

There exists a large variety of global optimization methods. They can be classified into deterministic and stochastic ones. Among the first ones are gradient-based methods, than can be applied when the cost function has adequate analytical properties. However, if the cost function is not smooth enough, it results difficult or impossible to apply these algorithms, and stochastic methods (such as Monte Carlo based ones) are more convenient. Moreover, a heuristic can be incorporated to the optimization algorithm to decide the next candidate to be tested or the way to compute the new candidate. Metaheuristic global optimization algorithms are those ones proposed to solve a general class of problems by using a combination of the cost function values and certain abstract reasoning rules, without paying attention to the specific nature of the problem. Sometimes, this combination is carried out in a stochastic way, either by considering samples in the search space or by using somehow randomness to obtain the optimal solution. A clear example of a metaheuristic stochastic global optimization algorithm is the standard simulated annealing (SA) method, in which the decision of the next candidate to be considered depends on the Boltzman probability distribution, as it will be described later in this paper. Other important examples of stochastic metaheuristic methods are genetic, swarm intelligence, parallel tempering and grenade explosion algorithms. Recently, metaheuristic algorithms have gained increasing scientific attention.

In this work, we focus on SA algorithm and its efficient parallelization on GPUs, which will lead us to use optimization algorithms that can also be understood as a kind of hybrid ones, combining SA and genetic algorithms (GA) (see {\cite{StornPrice-97}). They mainly consist of SA/GA with simple deterministic crossover operations (see \cite{Chen-98,ChenLeeParkMendes-2007,DeJohnThesis-75}).

SA is a metaheuristic stochastic optimization method that formulates the problem of finding the optimum of a cost function as the equilibrium configuration for a statistical thermodynamical system (see \cite{BM-95,DekkersAarts-91,Laarhoven-1987}). For a fixed temperature level, it has been first introduced by Metropolis et al. in \cite{Metropolis-1953}. Next, SA has been extended to the case of several temperatures, emulating the annealing process of steel forming, by Kirkpatrick et al. in \cite{Kirkpatrick-1983}.

Due to the great computational cost of SA, its parallelization has been analyzed by several authors and using different hardware architectures along time.
In   \cite{LeeLee-1996} Lee et al. studied different parallelization techniques based on the multiple Markov chains framework. Also several authors have analyzed different approaches in a SIMD (Single Instruction, Multiple Data) machine \cite{Chen-98}, depending on the number of communications performed between the independent Markov chains, and ranging from asynchronous to synchronous schemes with different periodicity in the communications. Special attention has been addressed in reducing the number of communications between processing threads, due the high latency of the communication network. In \cite{Rabenseifner-2005} a hybrid OpenMP/MPI implementation has been developed.

The parallelization of hybrid SA/GA algorithms has been analyzed by Chen et al. in \cite{Chen-98}. Moreover, in \cite{ChenLeeParkMendes-2007} a parallel hybrid SA/GA in MIMD PC clusters has been implemented, analyzing different crossover operations for generating the species.

Nowadays GPUs have become a cheap alternative to parallelize algorithms. The main objective of the present work is to develop a generic and highly optimized version of a SA algorithm for Nvidia GPUs in CUDA \cite{{NVIDIA-2011}}. For this purpose, first the more efficient versions
of SA presented in \cite{BM-95,Laarhoven-1987} have been analyzed, tested and  adapted to the GPU technology.

In the first section we present an introduction to the sequential simulated annealing algorithm.
Next we present the alternatives for the parallelization of the algorithm following the Multiple Markov chain approach. First a naive asynchronous implementation and then a synchronous implementation following \cite{LeeLee-1996} with communication between Markov chains at each temperature level is detailed. In GPUs decreasing the periodicity of the communications does not give a relevant difference in performance, because of the very low latency communication network between computing cores.

In the following section the precision of the algorithm is studied. Several classical optimization tests have been analyzed. A numerical convergence analysis is performed by comparing the sequential and parallel algorithms. Next the speedup of the parallel algorithm is studied attending to the different parameters of the optimization function and SA configurations.

Usually, the SA algorithm is used to obtain a starting point for a local optimization algorithm.
In this work we also present some examples of the precision and computational time, using our CUDA SA implementation and a Nelder-Mead algorithm.

\section{Simulated annealing}

\subsection{Sequential simulated annealing} \label{sec:seqSA}

As indicated in the introduction, SA is a stochastic optimization method which is mainly based on some statistical mechanics concepts. Thus, it formulates the problem of finding the optimum of a cost function in terms of obtaining the equilibrium configuration for a statistical thermodynamical system. Statistical mechanics is based on the description of a physical system by means of a set representing all possible system configurations and the probabilities of achieving each configuration. Thus, each set is associated with a partition function.

We say that a system is in equilibrium if the transition probability from state $S_i$ to state $S_j$, $P(S_i\to S_j)$, is the same as the probability of going from state $S_j$ to state $S_i$, $P(S_j\to S_i)$. A sufficient condition for equilibrium is the so called {\it detailed balance} or {\it local balance} condition, that can be written using the Bayesian properties:
\begin{equation}
\pi_i P(S_i\to S_j)=\pi_{j}P(S_j\to S_i),\label{eq:detailed-balance}
\end{equation}
where $\pi_i$ and $\pi_j$
are the probabilities of being in the states $i$ and $j$, respectively. These conditions can be also formulated in terms of Markov chains. A Markov process is said to be reversible (or time reversible), if it has a detailed balance
where $P(S_i\to S_j)$ denotes the Markov transition probability between the states $i$ and $j$.
That is, the forward and backwards Markov chains have the same transition probabilities.

Metropolis et al. proposed in \cite{Metropolis-1953} an algorithm for the simulation of atoms in equilibrium at a given fixed temperature.
It was based on the notion of detailed balance that describes equilibrium for
thermodynamical systems, whose configurations have probability proportional to the Boltzmann
factor.
The algorithm finds the transition probabilities for a Markov chain that yields the desired steady state distribution.
They introduced a random walk  (Markov chain of configurations) through the configuration space, using a fictitious kinetics. In this Markov chain approach, the time refers to the number of iterations of the  procedure. Moreover, we assume that our statistical system is considered to be in equilibrium so that the time results to be irrelevant. Starting from a set of transition probabilities,
a new set of transition probabilities satisfying the detailed balance condition can be found. This can be done by only accepting some of the transitions (see \cite{Bolstad-2001}). By appropriately using this procedure, the Markov chain converges to the steady state equilibrium distribution.

We aim to sample the space of possible configurations using a statistical thermodynamical system, that is in a thermal way. So, we force this system to satisfy the equation (\ref{eq:detailed-balance}). For the distribution function we chose the Boltzmann one, with degeneracy factor $1$, i.e. without repeated arrangements; which indicates the way the particles are distributed among the energy levels in a system in thermal equilibrium. More precisely, in order to define the probability of being at state $S_i$ at temperature $T$ we choose
\begin{equation}
\pi (S_i, T)=\dfrac{1}{Z(T)} \exp \Bigl( -\dfrac{ E_i}{k_b T} \Bigr),
\label{eq:bolztmann}
\end{equation}
where $k_b$ is the Boltzmann constant, $E_i$ is the energy level at state $S_i$ and $Z$ is a normalization function, also referred as the partition function, which depends on the temperature $T$ in the form
$$
Z(T)=  \displaystyle \sum_{j=1}^{N} \exp \left(-\frac{E_j}{k_bT} \right),
$$
where $N$ is the length of the Markov chain. Moreover, if the probability is given in terms of the Boltzmann distribution (\ref{eq:bolztmann}) then we have
\begin{equation}\label{eq:cociente_detailedbalance}
\dfrac{P(S_i \to S_j)}{P(S_j \to S_i)}=\dfrac{\pi (S_i, T)}{\pi(S_j, T)}=
\dfrac{\dfrac{1}{Z(T)} \exp \Bigl( -\dfrac{ E_i}{k_b T} \Bigr)}{\dfrac{1}{Z(T)}\exp \Bigl( -\dfrac{ E_j}{k_b T} \Bigr)}=
\exp \Bigl( -\dfrac{\Delta E_{ij}}{k_b T} \Bigr),
\end{equation}
with $\Delta E_{ij}=E_i-E_j$. Thus, the ratio in (\ref{eq:cociente_detailedbalance}) does not depend on $Z$. 

In \cite{Metropolis-1953} Metropolis et al. introduced a sufficient condition for the system to satisfy the detailed balance property. More precisely, the authors noticed that the
relative probability of equation (\ref{eq:cociente_detailedbalance}) could be obtained at simulation level by choosing
\begin{equation} \label{proba}
\dfrac{P(S_i \to S_j)}{P(S_j \to S_i)}=\begin{cases}
\exp \left( \dfrac{- \Delta E_{ij}}{k_b T}\right) & \text{if \ \ }  \Delta E_{ij} \geq 0, \\
\\
1 & \text{if \ \ } \Delta E_{ij} < 0.
\end{cases}
\end{equation}

By using the above election, the Markov chain satisfies the detailed balance condition. Therefore, if the trial satisfies condition (\ref{proba}) for the Boltzmann probability then the new configuration is accepted. Otherwise, it is rejected and the system remains unchanged. By using the appropriate physical units for energy and temperature we can take $k_b=1$, so that this strategy can be summarized as follows:

\begin{enumerate}
\item Starting from a configuration $S_i$, with known energy $E_i$, make a change in the configuration to obtain a new (neighbor) configuration $S_j$.

\item Compute $E_j$ (usually, it will be close to $E_i$, at least near the limit).

\item If $E_j < E_i$ then assume the new configuration, since it has lower energy (a desirable
property, according to the Boltzmann factor).

\item If $E_j \geq E_i$ then accept with probability $\exp(- \Delta E_{ij}/T)$ the new configuration (with higher energy) . This strategy implies that even when the temperature is higher in the new configuration, we don't mind following
steps in the {\em perhaps wrong} direction. Nevertheless, at lower temperatures we are more forced to
accept the lowest configuration we can find in our neighborhood and a jump to another region
is more unlikely to happen.
\end{enumerate}

Note that the original Metropolis algorithm is designed to find the optimum
configuration of the system at a fixed temperature.
Later on, the Metropolis algorithm has been generalized by Kirkpatrick et al. in \cite{Kirkpatrick-1983}, where an annealing schedule is introduced by defining how the temperature can be reduced. The algorithm starts with a high enough initial temperature, $T_0$, and the temperature is slowly decreased
by following a geometric progression, that is $T_n=\rho T_{n-1}$ with $\rho
<1$ (usually $0.9\le\rho<1$ to obtain a slow freezing procedure).  Thus, the SA algorithm consists of a temperature loop \cite{BM-95}, where the equilibrium state at each temperature is computed using the Metropolis algorithm. Therefore, the SA algorithm can be decomposed in the following steps (for example, see \cite{BM-95} for details):

\begin{itemize}
\item {\bf Step 1}: Start with the given temperature, $T_0$, and the initial point, $\pmb x_0$, with energy of configuration $E_0=f(\pmb x_0)$, where $f$ denotes the cost function of the problem (\ref{eq:min_pb}).

\item {\bf Step 2}:   Select a random coordinate of $\pmb x_0$ and a random number to modify the selected coordinate to obtain another  point $\pmb x_1 \in V$ in the neighborhood
of $\pmb x_0$.

\item {\bf Step  3}:  Compare  the  function  value at the two previous points, by using  the Metropolis criterion as  follows:  let $E_1=f(\pmb x_1)$ and select a sample, $u_1$, of a uniform random variable  ${\mathcal U}(0, 1)$. Then, move  the  system  to  the  new point if and
only if $u_1 <  \exp(- (E_{1}-E_{0})/T_0)$,
where $T_0$ is the current temperature.
In this way,  $E_1-E_0$ has been compared with  an  exponential  random variable with mean  $T_0$.  Note
that we  always move to the new point if $E_1  <  E_0$,  and  that  at  any  temperature there is  a  chance  for  the  system  to move ``upwards''. Note that we need three uniform random numbers: one to choose the coordinate, one to change the selected coordinate and the last one for the acceptance criterion.

\item {\bf Step 4}:  Either the  system has moved  or  not,  repeat steps  $2-3$.  At  each  stage we compare the  function  at the new  point  with  the  function at the previous point until the sequence of accepted points fulfills some test of achieving an equilibrium state.

\item {\bf Step 5}: Once the loop of the previous step has finished and an equilibrium state has been achieved for a given temperature, $T_0$, the temperature  is  decreased  according to  the  annealing  schedule, $T_1=\rho T_0$ (with a
decreasing factor $\rho$, $0<\rho<1$, usually $\rho$ close to one). Next, step $2$ starts again with temperature $T_1$ from the point obtained in the last iteration of the  algorithm as initial state. The iteration procedure continues until a stopping  criterion indicating that the system is enough frozen, in our case until achieving a target temperature $T_{min}$.

Notice that since we continue steps $2-3$ until an equilibrium  state, the starting values in
step $1$ have no effect on the solution. The algorithm can be implemented in numerous ways.

\end{itemize}

\subsection{Parallel Simulated Annealing}\label{sec:implementation}

The pseudocode of the algorithm described in the previous section can be sketched as follows:
\begin{center}
\framebox[0.95\linewidth]{
{\footnotesize
\begin{minipage}{\linewidth}
\begin{tabbing}
$\pmb x=\pmb x_0$; $T=T_0$;\\ %$\leftarrow$ GetInitialSolution()
do \= \\
\> for \=$j = 1$ to $N$ do\\
\>\>$\pmb x'=$ ComputeNeighbour($\pmb x$);\\
\>\>$\Delta E = f(\pmb x') - f(\pmb x)$;  \color{gray35} // \textit{Energy increment} \\
\>\>if \=$\big(\Delta E < 0$ or AcceptWithProbability $\exp(-\Delta E/T)\big)$\\
\>\>\> $\pmb x = \pmb x'$;  \color{gray35} // \textit{The trial is accepted} \\
\> end for\\
\>$T = \rho T$; \color{gray35} // with $0 < \rho < 1$\\
while ($T>T_{min}$);
\end{tabbing}
\end{minipage}
}
}
\end{center}

The SA algorithm is intrinsically sequential and thus it results difficult to
parallelize it without changing its recursive nature (see \cite{Chen-98}).

Several strategies can be followed in order to parallelize SA (see \cite{LeeLee-1996}, for example):

\begin{itemize}
\item {\it Application dependent parallelization}. The operations of the cost function are decomposed
among processors.
 \item {\it Domain decomposition}. The search space is sliced in several subdomains, each
processor searches the minimum at each subdomain and then shares its results with the rest of processors.
\item {\it Multiple Markov chains approach}. The most natural way to parallelize
SA is to follow a multiple Markov chain
strategy, where multiple Markov chains are executed asynchronously
and they communicate their final states every certain periods or at the end of
the process. This enables independent movements on each worker (SA chain) during
the intervals between consecutive communications. Attending to the number of
communications performed we can classify parallel implementation in different
categories.

The most straightforward approach is the case where the Markov chains
only communicate their states at the end of process. This is called asynchronous
approach (see \cite{KliewerTschoke-1998,LeeLee-1996,onbasoglu-2001}).

On the other hand, in the synchronous approach the Markov chains communicate
their states at intermediate temperature levels. Only function values are exchanged among
workers. The communication can be performed at each temperature level
(intensive exchange of solutions) or at a fixed number of temperature levels (periodic exchange of solutions).

Also we can classify the synchronous schemes attending to the type of the performed
exchange operation. This exchanged operation can be understood as a crossover genetic algorithm
operation where each Markov chain corresponds to an individual of a genetic algorithm. The
most simple crossover operation is taking the minimum among all the values returned by
the Markov chains at the current temperature level (see \cite{ChenLeeParkMendes-2007,LeeLee-1996,onbasoglu-2001}).

\end{itemize}

\subsubsection{Asynchronous} \label{sec:SAasincrono}

In order to parallelize SA, the most straightforward approach to take
advantage of the number of processors consists of simultaneously launching
a great number of SA processes. Thus, each processor performs a SA process asynchronously. At the final stage a reduce operation to obtain
the best optimum among all of the computed ones is performed. In this procedure,
either the initial configuration can be the same for all SA chains or a different starting
configuration for each processor can be randomly chosen (see Figure \ref{fig:sketchParaAsincrono}).

\begin{figure}[!htb]
\begin{center}
\includegraphics[height=4cm]{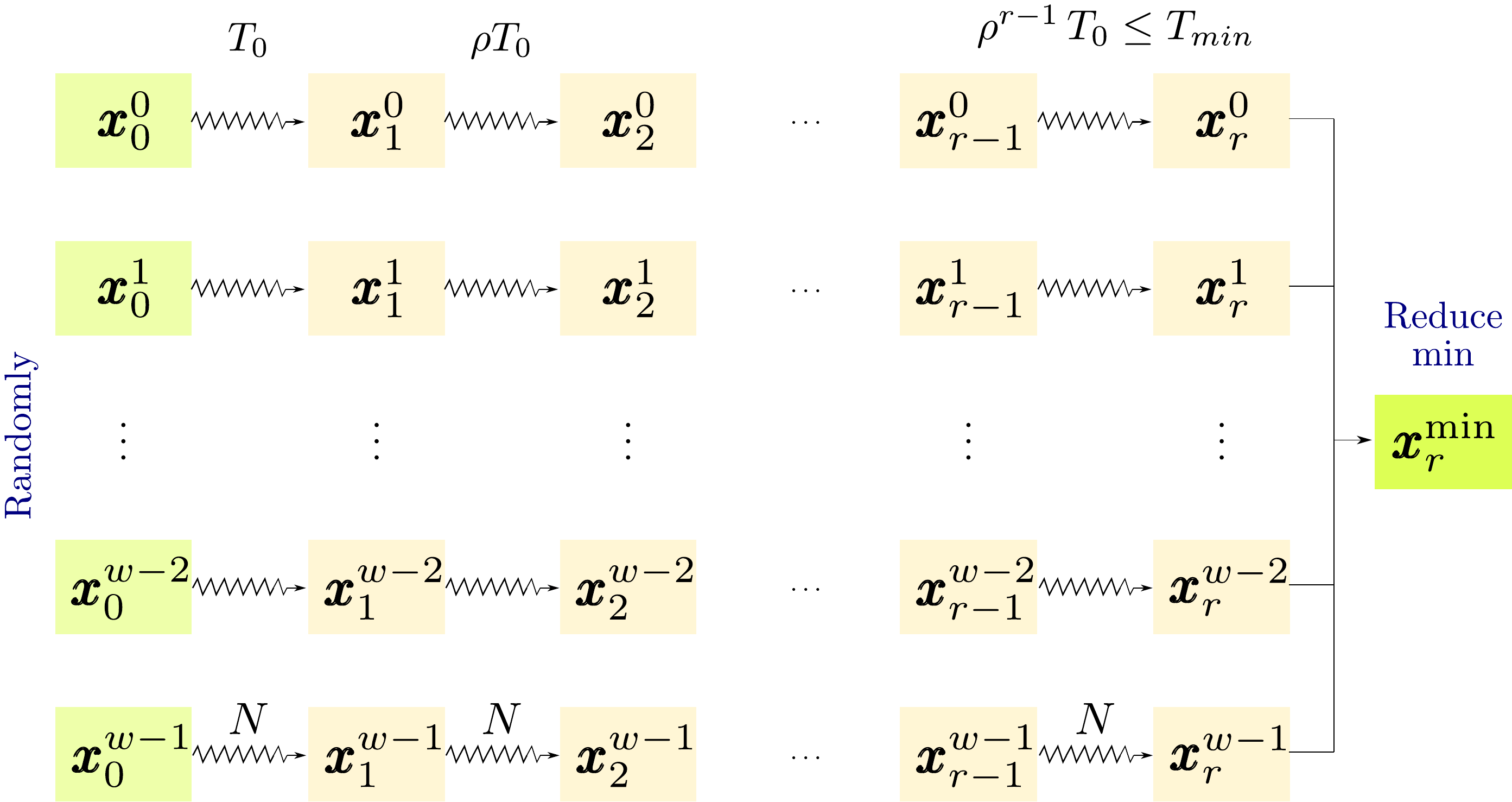}
\caption{Sketch of the asynchronous parallel algorithm.}
\label{fig:sketchParaAsincrono}
\end{center}
\end{figure}

\subsubsection{Synchronous approach with solution exchange at each temperature
level}

In the so called synchronous implementation, threads starts from a
random initial solution $\pmb x_0$, so that each thread runs
independently a Markov chain of constant length $N$ until reaching
the next level of temperature. As the temperature is fixed, each
thread actually performs a Metropolis process. Once all threads have finished,
they report their corresponding final states $\pmb x^p$ and the value $f(\pmb x^p)$,
$p=0,\ldots, w-1$ (where $w$ denotes the number of threads).
Next, a reduce operation to obtain the minimum of the cost function is performed.
So, if the minimum is obtained at a particular thread $p^{\star}$ then $\pmb x^{p^{\star}}$
is used as starting point for all threads at the following temperature level (see Figure \ref{fig:sketchParaSincrono}).
In the case of two or more points with the same objective function value,
the algorithm selects one of them and this choice does not affect the final result.

\begin{figure}[!htb]
\begin{center}
\includegraphics[height=4cm]{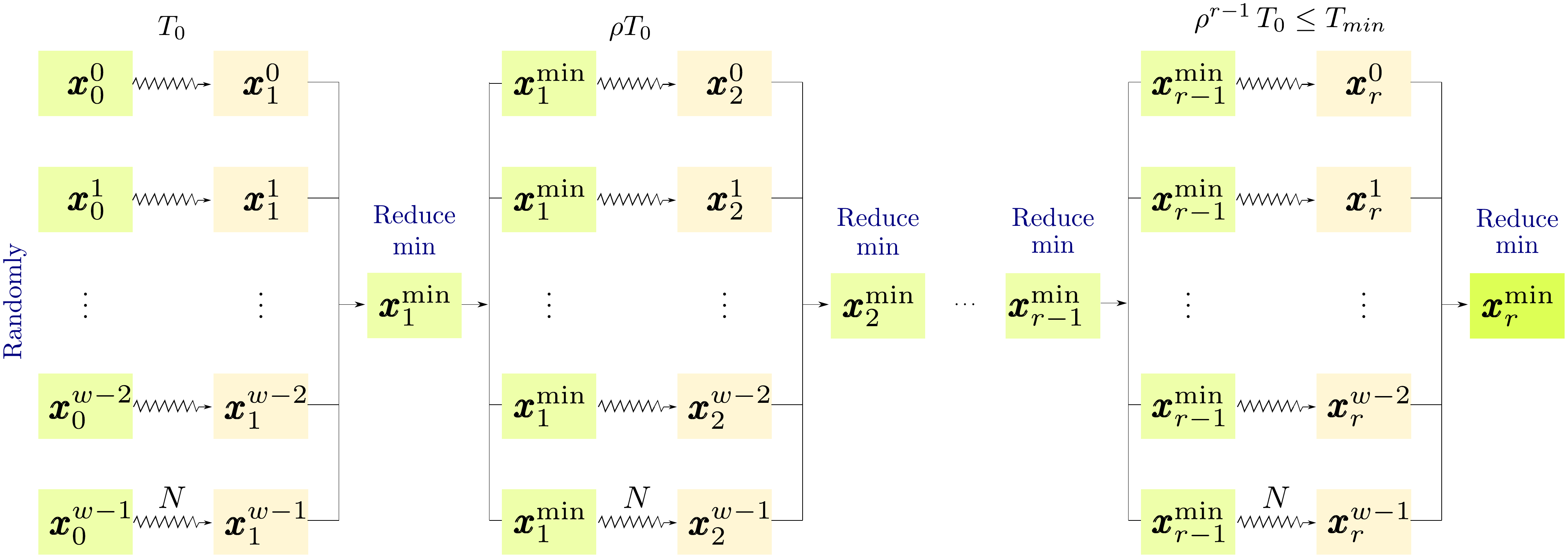}
\caption{Sketch of the synchronous parallel algorithm.}
\label{fig:sketchParaSincrono}
\end{center}
\end{figure}

This algorithm can be interpreted as a mixed technique of a genetic algorithm and a SA
one in which each independent Markov chain (SA process)
corresponds to a different individual in a genetic algorithm. Moreover, the
reduce operator can be understood as a crossover operation of a genetic
algorithm to select the evolution of these species. In \cite{LeeLee-1996} and
\cite{onbasoglu-2001} it is noted that for this algorithm the
independence of the Markov chains is lost: actually they depend on each other due to the use of a deterministic crossover operation (the minimum). This fact is overcome
in \cite{onbasoglu-2001} by introducing the so called {\it Synchronous approach with
occasional solution exchanges} (SOS) algorithm, where the authors propose a stochastic
crossover operator.

\section{Implementation on GPUs}

\subsection{General-Purpose Computing on Graphics Processing Units (GPGPU)}

From the mid nineties of 20th century, 3D capable Graphics Processing Units (GPUs), specialized graphics chips (coprocessors) independent from the CPU, started to be commonly used and integrated in computers. Pushed by the spectacular growth of graphics and videogames industries, always demanding more and more computing power, GPUs have spectacularly evolved during the last 10 years, becoming powerful and complex pieces of supercomputing hardware, with a massive parallel architecture.

Nowadays, a modern GPU consists of a many core processor, that can pack several hundreds (or even thousands) of computing cores/processors that work simultaneously and allows to execute many computing threads in parallel. Furthermore, all these cores can access to a common off-chip RAM memory by using a hardware topology that allows these threads to retrieve simultaneously several data from this memory, by performing memory access operations  also in parallel (under certain constraints). With all these processors working together, the GPU can execute many jobs in parallel. In Nvidia notation, we could call this architecture SIMT (Single Instruction, Multiple Threads), where a common program/piece-of-code (or computing kernel) is simultaneously executed by several threads over different data. This reminds the philosophy relying on the SIMD architecture.

As modern GPUs become more and more powerful in the last years, they increasingly attract the scientific community attention, which realized their potential to accelerate general-purpose scientific and engineering computing codes. This trend is called General-Purpose Computing on Graphics Processing Units (GPGPU), and consists of taking advantage of modern GPUs to perform general scientific computations.

Besides their intensive computational power, nowadays GPUs have become very popular in the supercomputing world, mainly because of the following advantages: they allow to save energy (as they are cheap and efficient in terms of Gflop per Watt), they are cheap (in terms of money per Gflop), and they also allow to save space (as many cores are packed into a small area).

As shown in the Top500 list (in June 2012), which lists the 500 more powerful supercomputers in the world (see \cite{TOP500}), three of the top ten supercomputers are heterogeneous systems, that use Nvidia  GPUs for offloading calculus.

However, GPUs  are very specialized and cannot live on their own. The GPU is a coprocessor that is used to accelerate applications running on the CPU, by offloading the most compute-intensive and time consuming portions of the code, although the rest of the application still runs on the CPU.  So, they depend on a CPU to control their execution.

Taking into account the large number of Markov chains that can be simultaneously computed to solve the minimization problem, the here treated algorithms are particularly well suited to be implemented in GPU technology.

Currently, there are two main GPU manufacturers, Nvidia and AMD (formerly ATI graphics).These two architectures are conceptually similar, although each one presents its own hardware peculiarities. In this work we have chosen Nvidia GPUs, the architecture of which is detailed in the next section.

\subsection{Nvidia GPUs, many core computing}

\begin{figure}[!htb]
\begin{center}
\includegraphics[width=12.5cm]{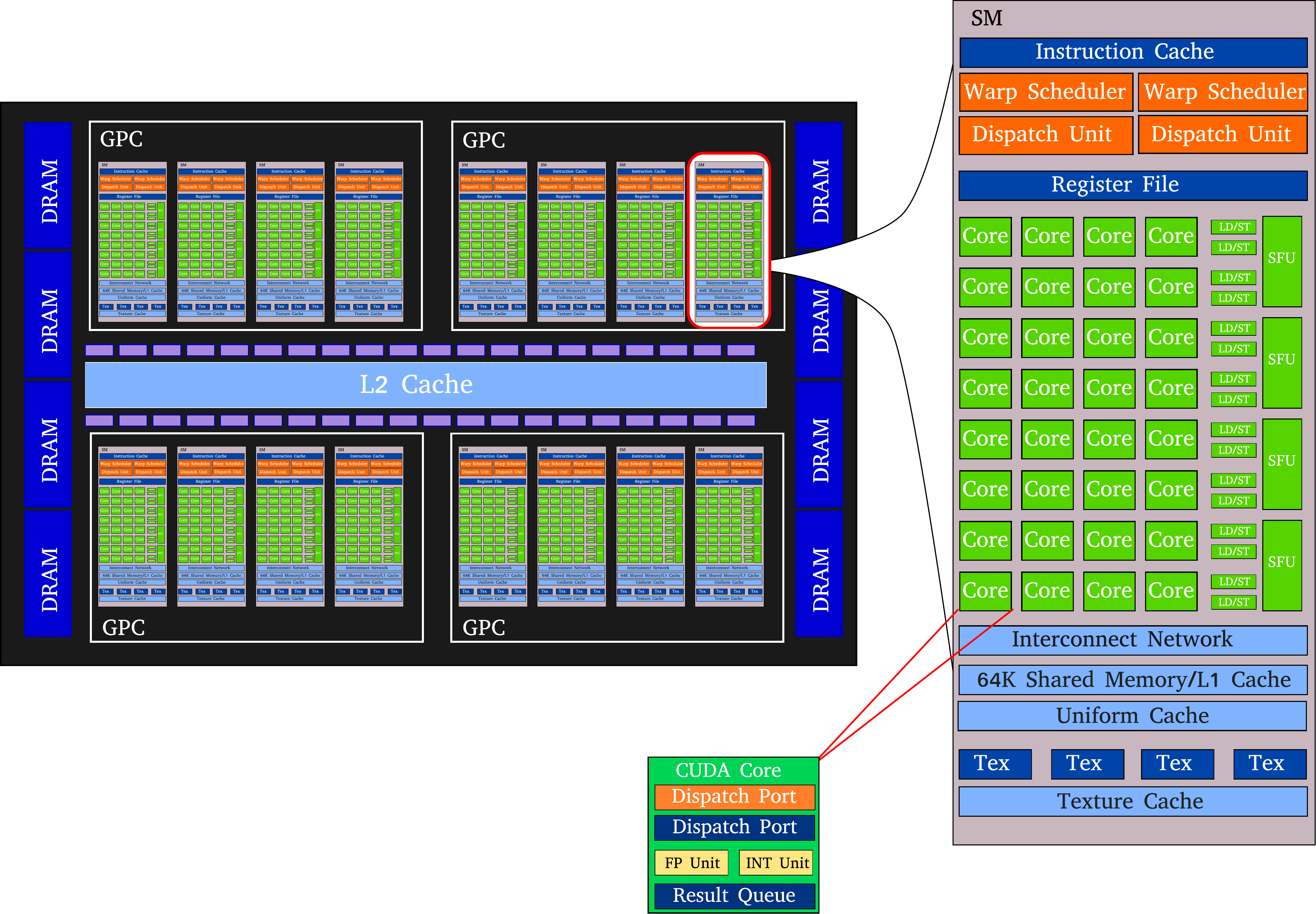}
\caption{Nvidia GPU hardware structure.}
\label{fig:gpu_sketch}
\end{center}
\end{figure}

A GPU (from now on, ``{\it the device}'') can be seen as a powerful SIMD coprocessor, endowed with a huge floating point processing power. Such coprocessor must be managed from a common CPU (from now on, ``{\it the host}''). In this work, we have used Nvidia GPUs, more precisely, the actual architecture, so called Fermi (or GF100), introduced in  early 2010. For a detailed explanation about this architecture, see \cite{fermi}.

As a physical layout (see Figure \ref{fig:gpu_sketch}), Nvidia Fermi GPUs chipsets are organized as a set of a variable number of Streaming Multiprocessors (SM) (from one to a maximum of sixteen, in the top Fermi models) grouped  into Graphics Processing Clusters (GPC). Each SM contains a variable number of  cores (or processors), $32$ in the case of the reference model of the GF100 chipsets. Each core contains a floating point unit and an integer unit. Each floating point unit can perform IEEE 754-2008 compliant double-precision floating point operations, in two clock cycles (half the performance of single-precision math).
The SM can process several execution threads at a time: they are planned and launched by a thread scheduler. The main differences between GPU and CPU threads are: firstly, GPU threads are extremely lightweight, i.e. very little creation overhead and instant switching; secondly, GPUs uses thousands of threads to achieve efficiency (instead, multi-core CPUs can use only a few).

Similarly to the CPU, Fermi GPUs have its own memory hierarchy:
\begin{itemize}
 \item {\it Device global memory}. The  GPU has its own high latency Random Access Memory (RAM) space (called device global memory), that is completely separate from the host memory. All transfers between these two memories have to be explicitly instructed by the programmer and these transfers have to be carefully designed because of the connection bandwidth (PCI Express $2.0$) and the memories latencies. Global memory space can be accessed at any time by all cores. The bandwidth from global memory to the SMs  is much bigger than the one of the CPU to its memory, with a peak of $\approx 200$ GB/s in top models. That is because under certain assumptions the global memory access by $32$ threads is coalesced into a single memory transaction, as soon as the data lies in the same segment of memory of size $128$ Bytes. On Fermi coalesced size is a full warp ($32$ threads). Also if data are not aligned we can achieve a high bandwidth using textures, specifically designed to exploit data spatial locality.

Previously to performing any calculus, the data to be processed must be pulled from the CPU to the GPU device memory and, once the calculations have finished, the computed results must be retrieved from the GPU.

Even with the device memory bandwidth being really high, it is not enough to  feed all the processors, i.e. to keep them fully occupied (note that all the processors have a theoretical peak performance of $520$ double-precision Gflops), so that a cache hierarchy is necessary.

\item A ``huge'' $768$ KB {\it L2 cache}. It is shared by all SMs and it manages the read/write and texture requests.

\item {\it Shared memory/L1 cache}. For low-latency access to data shared by cooperating threads in the same SM (implemented on chip).

Moreover, to benefit from frequently accessed data, each SM contains a low latency cache SRAM, referred as shared memory, of $64$ KB that is shared by all the cores of the SM, as a shared resource. In Fermi, this memory can be partitioned into a self-managed cache and a programmer-managed shared memory (in blocks of $48$ KB and $16$ KB).

\item {\it Texture cache}. With $12$ KB per SM, designed for small texture filtering operations, with spatial locality.

\item {\it Registers}. In addition to all these memories each SM contains a certain number of registers to store instruction operands (more precisely, in our case $32$ K registers of $32$-bits per multiprocessor).
\end{itemize}

From the programming point of view, similarly to the SIMD (Single Instruction, Multiple Data) execution model used for general data-parallel programming, the Nvidia model is SIMT (Single Instruction, Multiple Threads): the code execution unit is called a kernel and is executed simultaneously on all SMs by independent blocks of threads; each thread is assigned to a single processor and executes within its own execution environment (instruction address and register state), but they all run the same instruction at a time, although over different data. In order to efficiently execute hundreds of threads in parallel, the SM is hardware multithreaded. The SM also implements the synchronization  barrier with a single instruction. Once a block has its threads assigned to a SM, it is further divided by the SIMT multithreaded instruction unit into $32$-threads units called warps.

Each SM manages a pool of up to $48$ warps (giving a total of $1536$ threads), although only one of these warps will be actually executed by the hardware at any time instant. Threads are assigned to Streaming Multiprocessors in block granularity
(up to 8 blocks to each SM, for example $6$ blocks of $256$ threads). The size and number of blocks are defined by the programmer. Threads run concurrently and the SM maintains thread/block id's and  manages/schedules the threads execution.

There is an  API for programming  Nvidia GPUs called CUDA (Compute Unified Device Architecture) \cite{NVIDIA-2011},
which consists of: some drivers for the graphics card, a compiler and a language that is basically a set of extensions for the C/C++ language, that allows to control the GPU (the memory transfer  operations, the work assignment to the processors and the processors/threads synchronization).

CUDA provides all the means of a parallel programming model with the particularity of the previously cited types of memory.
There are CUDA instructions to manage all of these memories and to declare variables that are stored in any of those memory spaces. Inside the device, threads are able to access data from multiple memory spaces. Each thread block contains a shared memory which is visible to all threads of the block and with the same lifetime as the block. All threads from any grid have access to the same global memory which is persistent across kernel launches by the same application. Each transfer between these memory spaces must be also explicitly managed. CUDA also allows to work with texture memory to exploit data locality and with constant memory, used  to store small structures that are reused by all threads and that is also persistent across kernel launches. Transferring data between different types of memory inside the device results also important because of the different access patterns and the specific size and latency of the memory.

Thus due to the execution model and memory hierarchy, GPUs support two levels of parallelism:
\begin{itemize}
 \item An outer fully-parallel loop level that is supported at the grid level with no synchronization. Thread blocks in a grid are required to execute independently. It must be possible to execute them in any order, in parallel or in series. This independence requirement allows thread blocks to be scheduled in any order across any number of cores, thus enabling programmers to write scalable code.
 \item An inner synchronous loop level at the level of thread blocks where all threads within a block can cooperate and synchronize.
\end{itemize}

\subsection{Notes on the CUDA implementation}

In this section we detail the pseudocodes for the proposed asynchronous and synchronous versions of the parallel code.

The CUDA pseudocode of the asynchronous version is shown below:
\begin{lstlisting}[style=C,frame=single,label=list:SAasincrono,caption=Asynchronous simulated annealing.]
Initialize T = T_0
Initialize N, rho, T_min
Initialize n_blocks, n_threads_per_block
Initialize d_points = startPoint
Initialize bestPoint = 0

cusimann_kernel<<<n_blocks,n_threads_per_block>>>(T, N, rho, d_points, bestPoint)
bestPoint = reduceMin(d_points)
\end{lstlisting}
A kernel, \texttt{cusimann\_kernel}, that executes a sequential simulated annealing in each thread, is launched from the host (see the Listing \ref{list:SAasincrono}). The CUDA kernel is simultaneously executed in parallel by a large number of threads, thus allowing to compute a large number of Markov chains (in the here used GPU, GeForce GTX 470, the number of available CUDA cores is $448$). More precisely, this kernel launches a grid of \texttt{n\_blocks} thread blocks and each thread block groups \texttt{n\_threads\_per\_block} threads.

\begin{lstlisting}[style=C,frame=single,label=list:kernelSAasincrono,caption=Asynchronous simulated annealing kernel.]
__global__ void cusimann_kernel(T, N, rho, d_points, bestPoint) {
	
	Initialize global_tid
	Initialize x0 = d_points[global_tid] = bestPoint
	Initialize f_x0 = f(x0)
	do {
		for(i=0;i<N;i++){
			// Generate another point x1 in the neighborhood of x0
			d = Select randomly a coordinate of x0
			u = Select a random number to modify the selected d coordinate
			x1 = ComputeNeighbour(x0,d,u)
			f_x1 = f(x1)

			// If x1 satisfies the Metropolis criterion, move the system to x1
			if ( GenerateUniform(0,1) <= exp( -(f_x1-f_x0)/T ) )
				x0 = x1
				f_x0 = f_x1;
			end
	
		}
		T = T*rho
	} while (T>T_min)
}
\end{lstlisting}

\noindent Moreover, we take advantage of the constant memory to store the constant parameters, like $n$, $N$ and the box limits ($l_k$ and $u_k$), so that these data can be broadcasted to all threads. Furthermore, constant memory is cached, so that several consecutive accesses to the same memory position do not increase memory traffic. This is important because these consecutive accesses are repeatedly required by the SA algorithm.

As indicated in the Step 3 of the algorithm described in section \ref{sec:seqSA}, at each step of the Markov chain three uniform random numbers are required. At this point we take advantage of the Nvidia CURAND library \cite{NVIDIA-CURAND}, that allows parallel generation of random numbers to use them immediately by the kernels, without the extra time cost of writing and reading them from global memory.

As indicated in section \ref{sec:SAasincrono}, once all threads have finished to compute the Markov chains, the minimum of the function is obtained by a reduction operation (see Listing \ref{list:SAasincrono}). More precisely, for this purpose we need to find the index associated to the minimum of the vector storing the cost function values at the points returned by the threads. This operation is carried out in parallel inside the GPU, by means of the specific optimized Nvidia Thrust library, \cite{NVIDIA-THRUST}, that takes advantage of the coalesced memory access and the involved partial reductions are performed in shared memory.

Unlike the asynchronous version, in the synchronous one the temperature loop is carried out at CPU level, as detailed in the pseudocode in Listing \ref{list:SAsincrono},. Thus, at each temperature step the execution of the kernel detailed in Listing \ref{list:kernelSAsincrono}, as well as the reduction operation are required. As illustrated in the forthcoming Table \ref{tab:speed3vers}, the repeated use of the optimized reduction operation does not cause a significant performance overhead.

We also notice that in all implementations slow data transfers between CPU and global GPU memory do not appear.

\begin{lstlisting}[style=C,frame=single,label=list:SAsincrono,caption=Synchronous simulated annealing.]
Initialize T = T_0
Initialize N, rho, T_min
Initialize n_blocks, n_threads_per_block
Initialize d_points = startPoint
Initialize bestPoint = 0
do {
	cusimann_kernel<<<n_blocks,n_threads_per_block>>>(T, N, rho, d_points, bestPoint)
	bestPoint = reduceMin(d_points)
	T = T*rho
} while (T>T_min)
\end{lstlisting}

% The pseudocode for the kernel \texttt{cusimann\_kernel} is detailed in the Listing 4.

% \begin{program}
\begin{lstlisting}[style=C,frame=single,label=list:kernelSAsincrono,caption=Synchronous simulated annealing kernel.]
__global__ void cusimann_kernel(T, N, rho, d_points, bestPoint) {

	Initialize global_tid
	Initialize x0 = d_points[global_tid] = bestPoint
	Initialize f_x0 = f(x0)

	for(i=0;i<N;i++){
		// Generate another point x1 in the neighborhood of x0
		d = Select randomly a coordinate of x0
		u = Select a random number to modify the selected d coordinate
		x1 = ComputeNeighbour(x0,d,u)
		f_x1 = f(x1)

		// If x1 satisfies the Metropolis criterion, move the system to x1
		if ( GenerateUniform(0,1) <= exp( -(f_x1-f_x0)/T ) )
			x0 = x1
			f_x0 = f_x1;
		end

	}

}
\end{lstlisting}
% \end{program}

\section{Numerical experiments: academic tests}

In this section several experiments are presented to check the correctness and
performance of the here proposed CUDA implementation of SA. This CUDA implementation
has been developed from an optimized C code, following the ideas of section \ref{sec:implementation},
so that both codes perform exactly the same operations and their performance can thus be compared. The numerical experiments have been performed with the following hardware and software configurations: a GPU Nvidia Geforce GTX470,
a recent CPU Xeon E5620 clocked at 2.4 Ghz with 16 GB of RAM, CentOS Linux, NVIDIA CUDA SDK 3.2 and GNU C++ compiler 4.1.2.

In what follows, we denote by V0 the sequential implementation, by V1 the parallel asynchronous version and by V2 the parallel synchronous one.

\subsection{Analysis of a sample test problem: Normalized Schwefel function}

A typical benchmark for testing optimization techniques is the normalized Schwefel function (see \cite{ref:schwefel-func}, for example):
\begin{equation}
f(\pmb x) = -\dfrac{1}{n} \sum_{i=1}^{n} x_i \sin\left(\sqrt{|x_i|}\right), \quad -512 \leq x_i \leq 512, \quad \pmb x=(x_1,\ldots,x_n).
\end{equation}
% where $\pmb x\in \mathbb{R}^n$.
For any dimension $n$, the global minimum is achieved at the point $\pmb x^{\star}$, the coordinates of which are ${x^{\star}_i}=420.968746, \, i=1,\ldots,n$, and $ f({\pmb x^\star })=-418.982887$.

Table \ref{Tabla:N100} illustrates the accuracy for the three versions of the SA algorithm: sequential (V0), asynchronous (V1) and synchronous (V2). For these three versions we use the following configuration: $T_0=1000$, $T_{min}=0.01$, $N=100$ and $\rho=0.99$. For the parallel versions we use the choice $b=256$ and $g=64$, for the number of threads per block (block size) and the number of blocks per grid (grid size), respectively, so that the number of Markov chains is $16384$. With this configuration, the algorithm performs $1.8776\times 10^{9}$ function evaluations in all cases.

In order to take into account the impact of the random number seeds, we execute each algorithm $30$ times in all performed minimization examples. The synchronous version provides much better convergence results than the other two ones. Note that we have chosen the same SA configuration for all executions with different values of $n$ ($n=8, 16, 32, 64, 128, 256, 512$), so that the error obviously increases with the value of $n$. When the size of the problem increases, we should have selected a more restrictive SA setting because the minimization problem becomes more complex. Nevertheless, in order to compare in Table \ref{tab:speed3vers} the speedups of the parallel versions for different values of $n$ it results more convenient to consider the same SA setting for all cases.

Table \ref{tab:speed3vers} shows the performance of the GPU implementation with respect to a one core CPU. When $n$ increases the algorithm needs larger memory transfers, so that the speedup decreases. Notice that even for moderate values of $n$ the execution of the SA algorithm becomes memory-bounded, which means that its performance is limited by the memory bandwidth and not by the floating point performance.

In short, Tables \ref{Tabla:N100} and \ref{tab:speed3vers} show that the asynchronous version results to be a bit faster than the synchronous one, this is mainly because it does not perform reduction operations. Nevertheless, the errors are much larger in the asynchronous version. Notice that in all presented tables the computational time is expressed in seconds.

\begin{table}
\centering
{\tiny
\begin{tabular}{|r|c|c|c|c|c|c|}
\hline
 {$n$} &  \multicolumn{2}{c|}{V0} &  \multicolumn{2}{c|}{V1} &
\multicolumn{2}{c|}{V2}  \\
\hline
\, &$|f_{a}-f_{r}|$ & Relative error & $|f_{a}-f_{r}|$& Relative error
&$|f_{a}-f_{r}|$ & Relative error \\
\hline
$8$&  $1.3190\times 10^{-1}$ & $2.4283\times 10^{-3}$ & $1.2891\times 10^{-2}$ & $7.4675\times 10^{-4}$   &
$1.7000\times 10^{-5}$ &$4.1656 \times 10^{-5}$
\\
\hline
$16$& $2.3712\times 10^{-1}$ & $3.2557\times 10^{-3}$ & $7.4586\times 10^{-2}$  &   $1.8240\times 10^{-3}$  & $1.9000\times 10^{-6}$&
$5.0686\times 10^{-5}$ \\
\hline
$32$&  $3.3774\times 10^{-1}$ & $3.8852\times 10^{-3}$ & $2.8171\times 10^{-1}$ &  $3.5468\times 10^{-3}$ & $1.5730\times 10^{-4}$  &
$6.0577\times 10^{-5}$\\
\hline
$64$&   $7.9651\times 10^{-1}$ & $5.9664\times 10^{-3}$  & $9.7831\times 10^{-1}$ &  $6.6126\times 10^{-3}$ & $3.1880\times 10^{-4}$ &
$1.2132\times 10^{-4}$ \\
\hline
$128$&  $1.9198$ & $9.2648\times 10^{-3}$ & $3.0461$ &  $1.1674\times 10^{-2}$ & $1.2225\times 10^{-4}$
&$1.5304\times 10^{-4}$\\
\hline
$256$& $3.6230$  & $1.2733\times 10^{-2}$ & $9.5765$ & $8.0283\times 10^{-2}$  &   $1.4953\times 10^{-2}$ &
$8.2214\times 10^{-4}$\\
\hline
$512$& $7.3054$  & $1.8097\times 10^{-2}$ & $26.2282$&  $4.0424\times 10^{-1}$ & $4.6350\times 10^{-1}$
&$4.5503\times 10^{-3}$\\
\hline
\end{tabular}
}
\caption{Error of the solution obtained by the algorithm, both in the value of the function at the minimum (columns $|f_{a}-f_{r}|$, where $f_a$ is the objective function value found by the algorithm and $f_r$ is the exact function value in the real minimum) and in the minimum (columns Relative error, measured in $\parallel \cdot \parallel_2$).}
\label{Tabla:N100}
\end{table}

\begin{table}[!htb]
\centering
{\footnotesize
\begin{tabular}{|l||r||r|r||r|r|}
\hline
$n$   & {V0}  & \multicolumn{2}{c||}{{V1}}  & \multicolumn{2}{c|}{{V2}}\\
\hline
 & Time & Time   & Speedup  & Time  & Speedup   \\
\hline
$8$  & $1493.7686$ & $5.5436$ & $269.4595$ & $5.6859$ & $262.7121$\\
\hline
$16$  & $2529.3072$ &$15.3942$ & $164.3027$  &  $15.5889$& $162.2502$
\\
\hline
$32$ & $4618.5820$ &$56.9808$ &  $81.0550$ & $60.1882$ & $76.7356$   \\
\hline
$64$ & $8773.0560$ &$106.6075$ & $82.2930$  & $110.2702$ & $79.5596$
\\
\hline
$128$ & $17169.0000$ &$210.9499$ &  $81.3890$  & $215.5416$& $79.6552$\\
\hline
$256$ & $34251.9240$ & $455.4910$ &  $75.1978$ & $462.8035$& $74.0096$  \\
\hline
$512$ & $68134.5760$ & $871.7434$ &  $78.1589$ & $893.7668$& $76.2330$  \\
\hline
\end{tabular}
\caption{\footnotesize Performance of CUDA version vs. sequential version with
one CPU core for different number of parameters.} \label{tab:speed3vers}
}
\end{table}

\subsubsection{Numerical convergence analysis}

In this case we compare the convergence of the two parallel algorithms with the
sequential version. For this purpose the same number
of explored points in the function domain are considered. In Figure \ref{fig:grafsConv} three graphics of the relative error vs. the number of explored points for $n=8, 16$ and $32$ are presented. In Figure \ref{fig:grafsConv2} the same graphics for $n=64, 128$ and $256$ are shown. From them, it is clear that the synchronous version converges more quickly. In order to compare the asynchronous version with the other two ones at a given temperature step of SA, we must choose a point that summarizes the state where the different threads are. For this purpose we have chosen the best point of all threads, so that we are very optimistic in representing the convergence of the asynchronous version.

\begin{figure}[!htbp]
\begin{center}
\includegraphics[width=10.5cm]{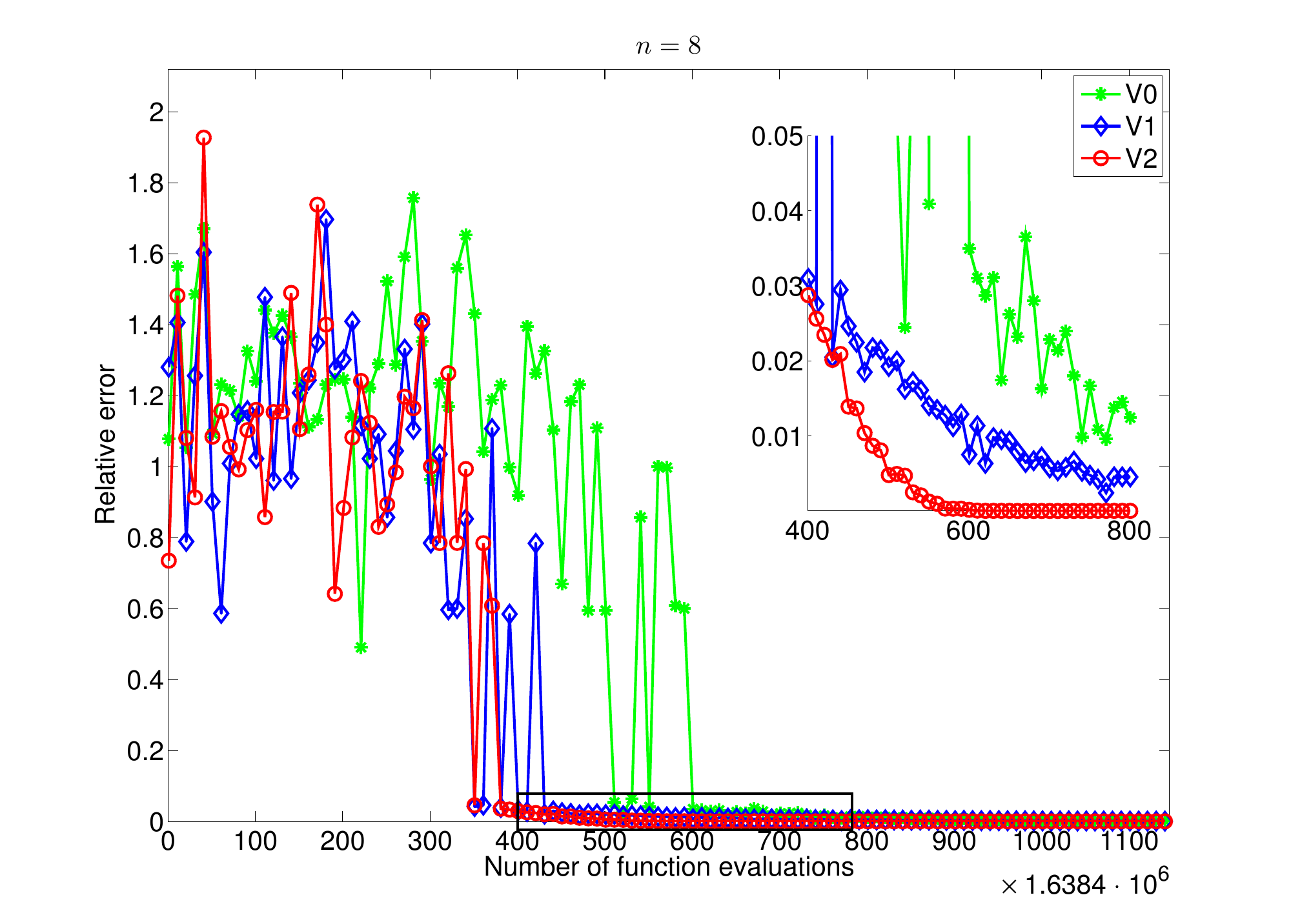}

\includegraphics[width=10.5cm]{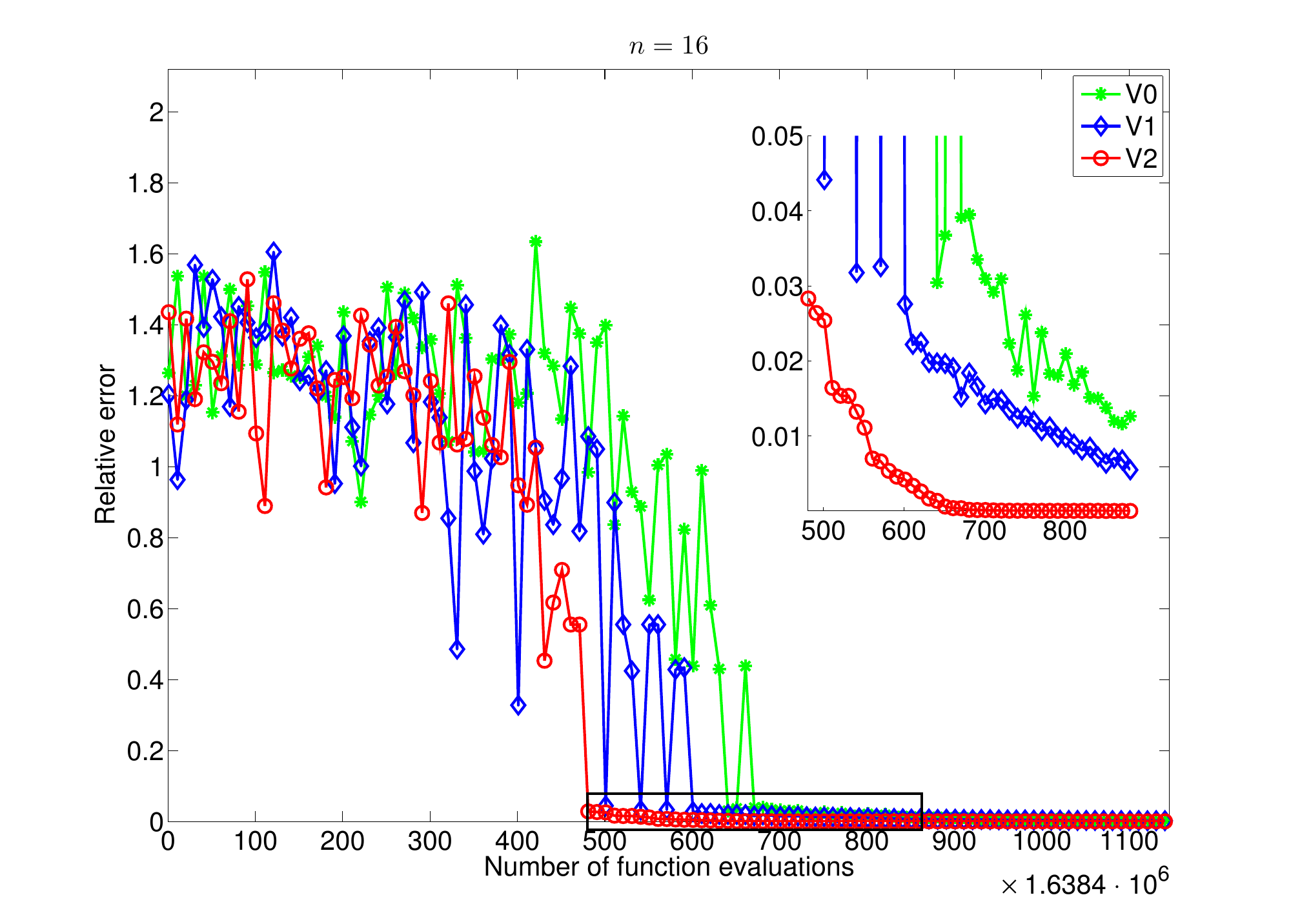}

\includegraphics[width=10.5cm]{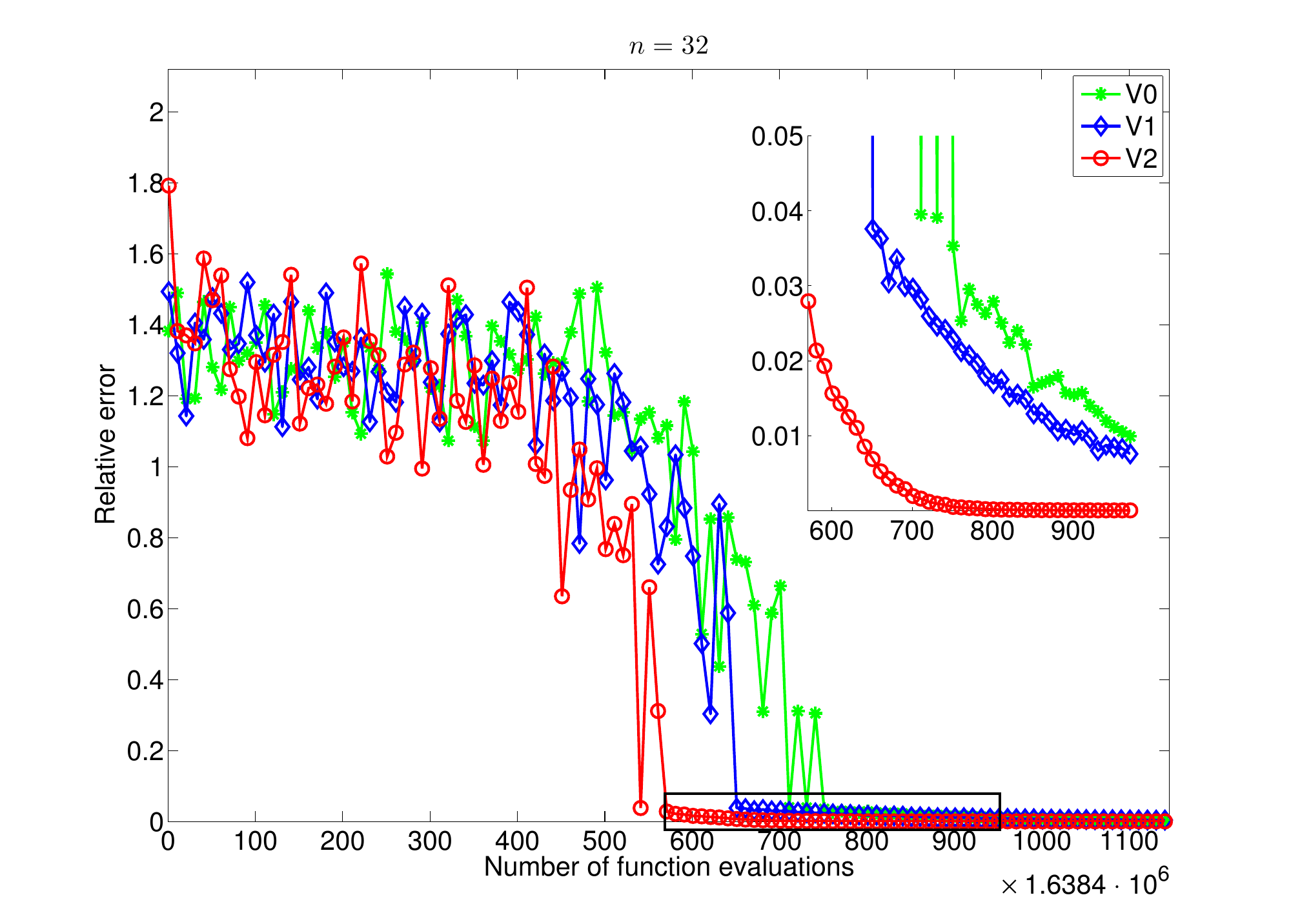}

\caption{For the three versions V0, V1 and V2, convergence rate for different runs with $n = 8, 16$ and $32$.}
\label{fig:grafsConv}
\end{center}
\end{figure}

\begin{figure}[!htbp]
\begin{center}
\includegraphics[width=10.5cm]{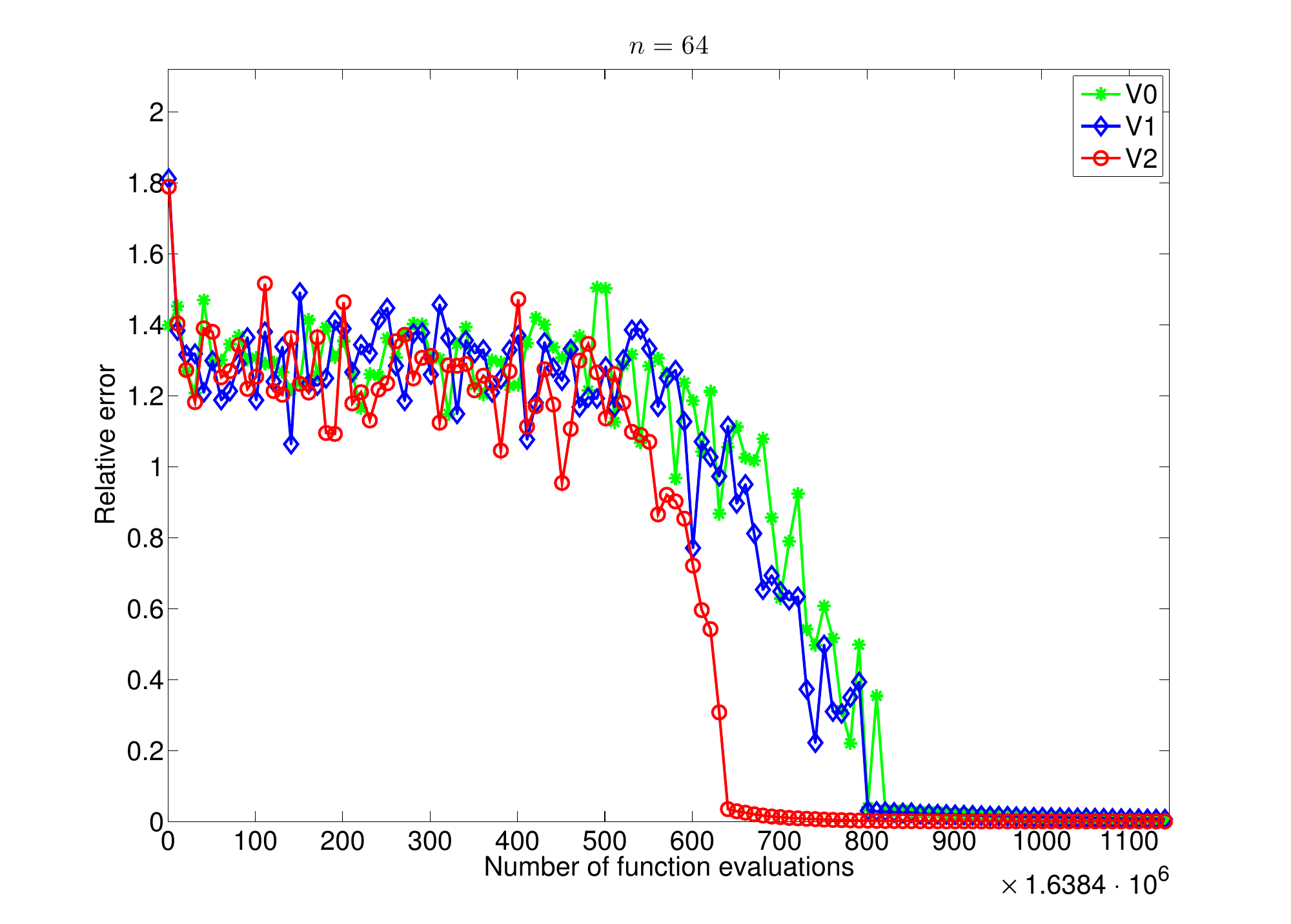}

\includegraphics[width=10.5cm]{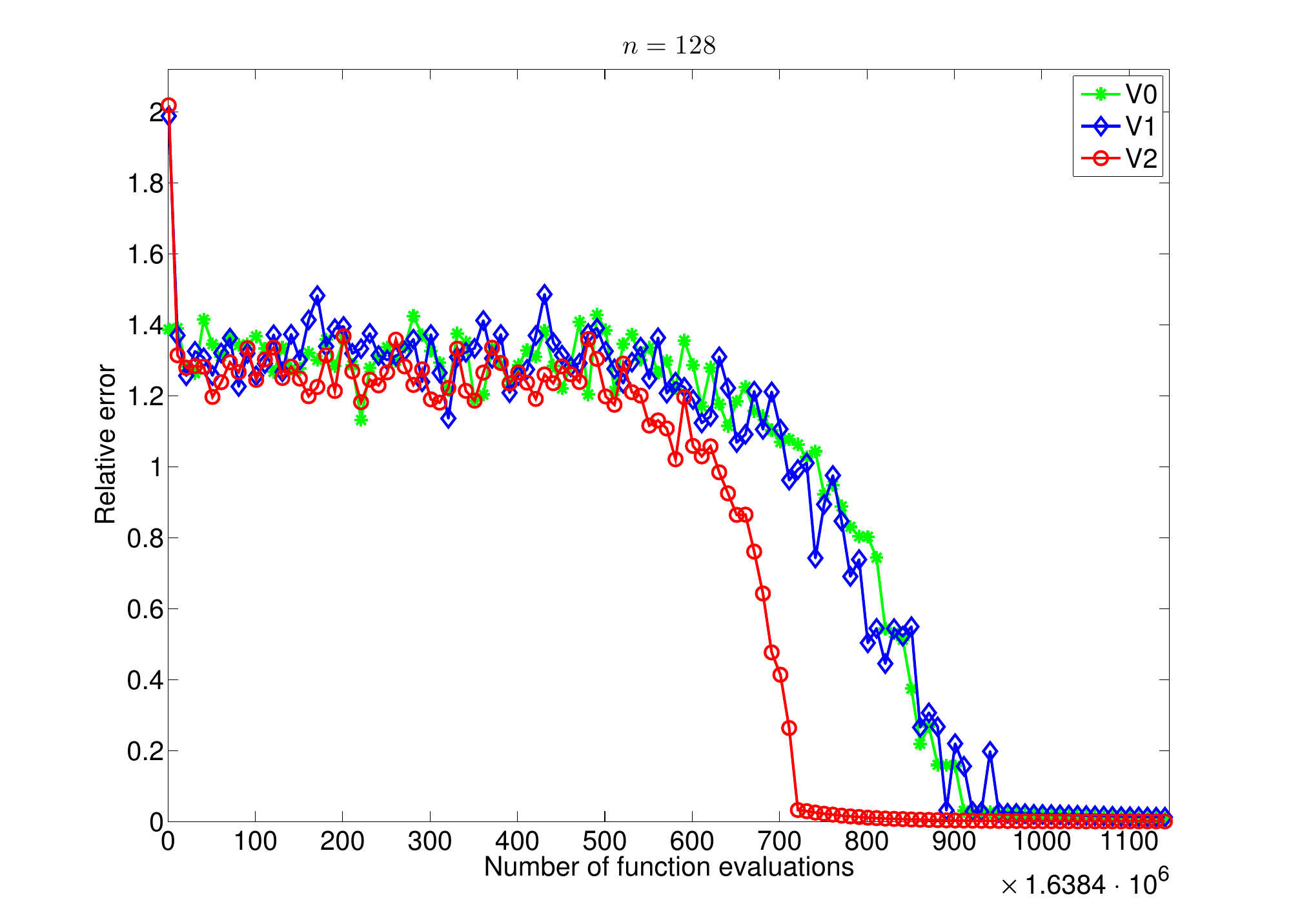}

\includegraphics[width=10.5cm]{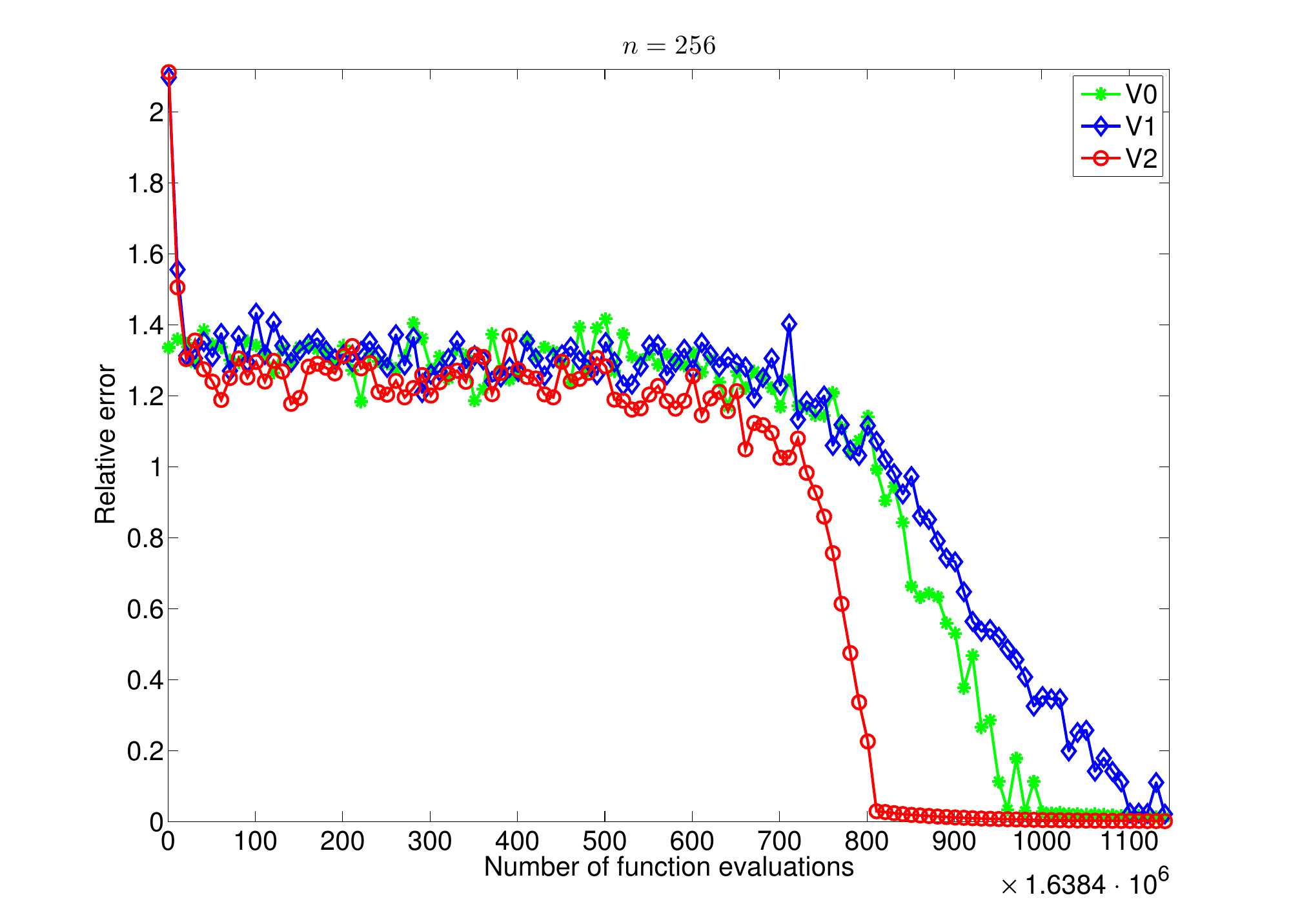}

\caption{For the three versions V0, V1 and V2, convergence rate for different runs with $n = 64, 128$ and $256$.}
\label{fig:grafsConv2}
\end{center}
\end{figure}

As expected, all the results presented so far show that the synchronous version results better to approximate the solution, specially for higher dimension problems. Therefore, in the forthcoming subsections we only analyze the behavior of this synchronous version.

\subsubsection{Increasing the number of launched threads}

At this point we analyze the algorithm behavior when increasing the number of launched threads. Table \ref{Tabla:errorVSthreads} illustrates how the error of the obtained solution is reduced when we successively multiply by $100$ the initial number of launched threads. Table \ref{Tabla:speedupVSthreads} shows how the speedup increases when we multiply by $2$ the initial number of launched threads, not only in the cases where the execution is not memory-bounded ($n=16$), but also in the cases where it is memory-bounded ($n=32$). Note that even in the memory-bounded case the obtained speedup is around $90$.

\begin{table}
\centering
{\footnotesize
\begin{tabular}{|r|r|r|c|}
\hline
 Threads & Function evaluations & $|f_{a}-f_{r}|$ & Relative error  \\
\hline
\hline
$768$ & $2.7648\times 10^{4} $ & $47.7821$ & $1.1085$  \\ % 0.037560 segundos la secuencial
\hline
$76800$ & $ 2.7648\times 10^{6}$ & $8.0830$ & $1.9117\times 10^{-2}$  \\ % 3.741118 segundos la secuencial
\hline
$7680000$ & $2.7648\times 10^{8} $ & $1.4345$ & $8.0156\times 10^{-3}$  \\ %373.3963 segundos la secuencial
\hline
\end{tabular}
}
\caption{\footnotesize Behavior of the errors when increasing the number of launched threads. Tests were performed with $n=16$, $T_0=5$, $T_{min}=0.5$, $\rho=0.7$, $N=5$.}
\label{Tabla:errorVSthreads}
\end{table}

\begin{table}
\centering
{\footnotesize
\begin{tabular}{|c|c||c|c||c|c|}
\hline
 Threads & Function evaluations & \multicolumn{2}{c||}{$n=16$} & \multicolumn{2}{c|}{$n=32$} \\
\hline
 &   & Time & Speedup & Time & Speedup \\
\hline
$128\times64$ & $9.3881\times 10^{8} $ & $10.5519$ & $122.1563$ & $31.8715 $ & $73.4358 $ \\
\hline
$256\times64$ & $1.8776\times 10^{9} $ & $15.5889$ & $162.2502$ & $ 57.0946$ & $ 81.9460$ \\
\hline
$256\times128$ & $3.7552\times 10^{9} $ & $25.4284$ & $203.3326$ & $109.0251 $ & $85.9172 $ \\
\hline
$256\times256$ & $7.5105\times 10^{9} $ & $46.4328$ & $222.1904$ & $211.5324 $ & $88.5479 $ \\
\hline
$256\times512$ & $1.5021\times 10^{10}$ & $87.7999$ & $235.5945$ & $ 414.3638$ & $ 90.5019$ \\
\hline
\end{tabular}
}
\caption{\footnotesize Behavior of the speedup when increasing the number of launched threads. Tests were performed with $T_0=1000$, $T_{min}=0.01$, $\rho=0.99$, $N=100$.}
\label{Tabla:speedupVSthreads}
\end{table}

\subsubsection{Increasing the length of Markov chains}

Table \ref{Tabla:speedupVSN} shows the behavior of the speedup when successively doubling $N$, which denotes the length of the Markov chains, also for both cases $n=16$ (not memory-bounded) and $n=32$ (memory-bounded). Notice that the speedups are maintained even for large lengths of Markov chains.

\begin{table}
\centering
{\footnotesize
\begin{tabular}{|c|c||c|c||c|c|}
\hline
 $N$ & Function evaluations & \multicolumn{2}{c||}{$n=16$} & \multicolumn{2}{c|}{$n=32$} \\
\hline
 &   & Time & Speedup & Time & Speedup \\
\hline
$50$ & $9.3881\times 10^{8}$ & $9.3158$ & $138.4039$ & $30.1144 $ & $77.9015 $  \\
\hline
$100$ & $1.8776\times 10^{9}$ & $15.5889$ & $162.2502$ & $57.0946 $ & $ 81.9460$ \\
\hline
$200$ & $3.7552\times 10^{9}$ & $28.4907$ & $ 181.4357$ & $ 111.2561$ & $ 84.2041$ \\
\hline
$400$ & $7.5104\times 10^{9}$ & $54.1433$ & $191.1686$ & $219.3096 $ & $85.6900 $  \\
\hline
$800$ & $1.5021\times 10^{10}$ & $105.4553$ & $196.1659$ & $ 435.2572$ & $86.1849$  \\
\hline
$1600$ & $ 3.0042\times 10^{10}$ & $208.1954$ & $ 198.1213$ & $869.2079 $ & $86.2000 $  \\
\hline
$3200$ & $6.0083\times 10^{10}$ & $413.3363$ & $199.5752 $ & $1732.5052 $ & $ 86.5688$  \\
\hline
\end{tabular}
}
\caption{\footnotesize Behavior of the speedup when increasing $N$. These tests were performed with the following configuration of simulated annealing, $T_0=1000$, $T_{min}=0.01$, $\rho=0.99$, $b=256$, $g=64$.}
\label{Tabla:speedupVSN}
\end{table}

\subsubsection{Increasing the number of function evaluations}
Table \ref{Tabla:speedupVSn} shows how the speedup evolves when we increase the number of function evaluations by approximately successively doubling the initial value, also in both cases $n=16$ (not memory-bounded) and $n=32$ (memory-bounded). In practice, the doubling of the number of function evaluations is achieved by different procedures: doubling the length of the Markov chain, doubling the number of launched threads, increasing the gap between the initial and the target minimum temperature or increasing the value of $\rho$.

\begin{table}
\centering
{\footnotesize
\begin{tabular}{|c||c|c||c|c|}
\hline
  Function evaluations & \multicolumn{2}{c||}{$n=16$} & \multicolumn{2}{c|}{$n=32$} \\
\hline
& Time & Speedup & Time & Speedup \\
\hline
$1.8776\times 10^{9}$ & $15.6681$ & $162.2502$ &  $60.1882$ & $76.7356$  \\
\hline
$3.7552\times 10^{9}$ & $25.4315$ & $203.0942$ &  $109.1611 $ & $85.6823 $  \\
\hline
$ 7.5105\times 10^{9}$ & $47.8059$ & $215.8074$ &  $ 215.0048$ & $87.0605 $  \\
\hline
$ 1.5021\times 10^{10}$ & $92.7941$ & $222.3886$ &  $ 426.6933 $ & $87.6947 $  \\
\hline
$ 3.0042\times 10^{10}$ & $182.6766$ & $225.6187$ & $850.9985 $ & $88.0599 $  \\
\hline
\end{tabular}
}
\caption{\footnotesize Behavior of the speedup when increasing the number of function evaluations. }
\label{Tabla:speedupVSn}
\end{table}

\subsubsection{Double vs. Float}

Table \ref{Tabla:precisionVStime} shows that executions in double-precision are twice slower than in single one. Notice that in the best scenarios for the HPC versions of the Fermi architecture the double precision speed results to be limited to one half of the single-precision one. Obviously, the obtained error with double-precision is lower, but single-precision accuracy is enough because the purpose of the SA algorithm is to find an approximate minimum (see \cite{Kirkpatrick-1983}). This is the reason why the results presented in all tables have been obtained with single-precision.

\begin{table}
\centering
{\footnotesize
\begin{tabular}{|c|c|c|}
\hline
  & Time & Relative error  \\
\hline
\hline
Single-precision & $15.5889$  & $5.0686\times 10^{-5}$ \\
\hline
Double-precision &  $32.3916$ & $2.1166\times 10^{-7} $  \\
\hline
\end{tabular}
}
\caption{\footnotesize Computational times in seconds and relative quadratic errors with single and double-precision for the next simulated annealing configuration: $n=16$, $T_0=1000$, $T_{min}=0.01$, $\rho=0.99$, $b=256$, $g=64$.}
\label{Tabla:precisionVStime}
\end{table}

\subsection{The set of performed tests}

 In the previous section a particular minimization problem has been deeply analyzed. Furthermore, the proposed CUDA implementation for SA algorithm in GPUs has been tested against a large enough number of appropriate examples. A brief description of the different optimization problems that have been considered in the benchmark is listed in the Appendix. The number and the kind of problems included in the Appendix are chosen so that they are enough to obtain some conclusions from them and the test suite should not be overwhelming so that this study is unmanageable. Finally, the suite contains $41$ examples. Table \ref{Table:number-test-functions} lists these problems with the corresponding number of variables of each one. Moreover, the comparative analysis of results mainly focuses on the objective function values and in the locations of the solutions in the domain space obtained by the SA algorithm. Table \ref{academico2} shows the obtained results, both for the asynchronous and synchronous versions. In Table \ref{academico2} SA configurations that achieve small errors in the synchronous version are considered. Therefore, execution times become high for functions with a large number of parameters or with a large number of local minima. Since typically in many real applications the hybrid approaches (in which SA provides a starting point for a local minimization algorithm) are widely used \cite{BM-95}, we present in Table \ref{Tabla:hybrid} the obtained results using a hybrid strategy with Nelder-Mead as local minimizer. Both execution times and errors are much smaller when appropriately combining the SA and the local minimization algorithm.

\begin{table}[!htb]
{\footnotesize
\centering
\begin{tabular}{|l|c|c|}
\hline
 Function $f$ reference & Name of the problem & Dimension $n$  \\
\hline
F0\_a & Schwefel problem  & 8 \\
F0\_b & Schwefel problem  & 16 \\
F0\_c & Schwefel problem  & 32 \\
F0\_d & Schwefel problem  & 64 \\
F0\_e & Schwefel problem  & 128 \\
F0\_f & Schwefel problem  & 256 \\
F0\_g & Schwefel problem  & 512 \\
F1\_a & Ackley problem  & 30 \\
F1\_b & Ackley problem  & 100 \\
F1\_c & Ackley problem  & 200 \\
F1\_d & Ackley problem  & 400 \\
F2 & Branin problem  & 2 \\
F3\_a & Cosine problem & 2 \\
F3\_b & Cosine problem & 4 \\
F4 & Dekkers and Aarts problem & 2 \\
F5 & Easom problem & 2 \\
F6 & Exponential problem & 4 \\
F7 & Goldstein and Price problem & 2 \\
F8\_a & Griewank problem & 100 \\
F8\_b & Griewank problem & 200 \\
F8\_c & Griewank problem & 400 \\
F9 & Himmelblau problem & 2 \\
F10\_a & Levy and Montalvo problem & 2 \\
F10\_b & Levy and Montalvo problem & 5 \\
F10\_c & Levy and Montalvo problem & 10 \\
F11\_a & Modified Langerman problem & 2 \\
F11\_b & Modified Langerman problem & 5 \\
F12\_a & Michalewicz problem & 2 \\
F12\_b & Michalewicz problem & 5 \\
F12\_c & Michalewicz problem & 10 \\

F13\_a & Rastrigin problem & 100 \\
F13\_b & Rastrigin problem & 400 \\
F14 & Generalized Rosenbrock problem & 4 \\

F15 & Salomon problem & 10 \\
F16 & Six-Hump Camel Back problem & 2 \\
F17 & Shubert problem & 2 \\
F18\_a & Shekel 5 problem, $m=5$ & 4 \\
F18\_b & Shekel 7 problem, $m=7$ & 4 \\
F18\_c & Shekel 10 problem, $m=10$ & 4 \\
F19\_a & Modified Shekel Foxholes problem & 2 \\
F19\_b & Modified Shekel Foxholes problem & 5 \\
\hline
\end{tabular}
\caption{\footnotesize Set of test problems, where first column indicates the assigned reference to display results.}
\label{Table:number-test-functions}
}
\end{table}

\begin{table}[!htb]
\centering
{\tiny
\begin{tabular}{|r||l||r|l|l||r|l|l|}
\hline
$f$   & Function eval.  & \multicolumn{3}{c||}{ V1}
& \multicolumn{3}{c|}{V2}\\
\hline
        &   & Time  &
$|f_{a}-f_{r}|$ & Error & Time &
$|f_{a}-f_{r}|$ & Error \\
\hline
F1\_a  & $2.25 \times 10^{9}$   & $56.03 $ & $8.84 \times 10^{-2}$ & $9.79\times 10^{-2}$ &  $56.42$ &
$3.20\times 10^{-5}$ &$4.56\times 10^{-5}$ \\
\hline
F1\_b  & $2.25\times 10^{9}$  & $221.36$ & $9.45\times 10^{-1}$ & $1.11$ &  $212.11$&
$1.69\times 10^{-4}$ & $4.26\times 10^{-4}$ \\
\hline
F1\_c  & $2.63\times 10^{11}$ & $52132.20$ & $ 1.52\times 10^{-1}$ & $3.95\times 10^{-1} $ & $52916.78$ & $1.93\times 10^{-4}$
& $6.89\times 10^{-4}$ \\
\hline
F1\_d  & $2.63\times 10^{11}$  & $100481.20 $ & $3.34\times 10^{-1}$  & $1.01$  & $101754.88$ & $3.90\times 10^{-4}$ & $1.95\times 10^{-3}$ \\
\hline
F2  & $1.50\times 10^{9}$  & $4.14$ & $1.00\times 10^{-7}$ & $9.90\times 10^{-4}$ & $4.22$& $1.00\times 10^{-7}$ & $1.09\times 10^{-4}$ \\
\hline
F3\_a  & $1.87\times 10^{9}$  & $4.18$ & $1.00\times 10^{-6}$ & $2.21\times 10^{-4}$ & $4.32$& $1.00\times 10^{-7}$ & $2.08\times 10^{-5}$ \\
\hline
F3\_b  & $1.87\times 10^{9}$ & $4.93$ & $1.02\times 10^{-4}$ & $3.01\times 10^{-3}$ & $5.05$& $1.00\times 10^{-7}$ & $3.63\times 10^{-5}$ \\
\hline
F4  & $1.87\times 10^{9}$  & $4.38$ & $3.20\times 10^{-3}$ & $2.11\times 10^{-5}$ & $4.53$& $4.20\times 10^{-4}$ & $1.92\times 10^{-5}$
\\
\hline
F5  & $1.87\times 10^{9}$  & $4.19$ & $3.00\times 10^{-6}$ & $2.54\times 10^{-4}$ & $4.32$& $1.00\times 10^{-7}$ & $4.15\times 10^{-5}$ \\
\hline
F6  & $2.25\times 10^{9}$  & $4.39$ & $9.00\times 10^{-6}$ & $4.20\times 10^{-3}$ & $4.63$& $1.00\times 10^{-7}$ & $3.58\times 10^{-4}$ \\
\hline
F7  & $1.87\times 10^{9}$  & $4.13$ & $3.00\times 10^{-5} $ & $1.16\times 10^{-4}$ & $4.15$& $2.70\times 10^{-5} $
&$3.46\times 10^{-5}$ \\
\hline
F8\_a  & $3.37\times 10^{9}$  & $674.93$ & $1.11$ & $2.16\times 10^{1}$ & $666.02$& $1.00\times 10^{-7}$ & $2.80\times 10^{-3}$ \\
\hline
F8\_b  & $5.25\times 10^{9}$ & $2102.46 $ & $1.59 $ & $4.87\times 10^{1}$ & $2090.60$ & $3.00\times 10^{-6}$ & $2.69\times 10^{-2}$ \\
\hline
F8\_c  & $3.37\times 10^{9}$ & $2586.84$ & $4.27$ & $1.14\times 10^{2}$ & $2536.51$ & $5.43\times 10^{-3}$ & $1.40$ \\
\hline
F9  & $1.87\times 10^{9}$  & $3.90$ & $2.00\times 10^{-6}$ & $5.06\times 10^{-5}$ & $4.08$& $1.00\times 10^{-7}$ & $1.00\times 10^{-7}$ \\
\hline
F10\_a  & $2.25\times 10^{9}$  & $4.77$ & $1.00\times 10^{-7}$ & $4.45\times 10^{-4}$ & $4.93$& $1.00\times 10^{-7}$ & $3.28\times 10^{-7}$ \\
\hline
F10\_b  & $2.25\times 10^{9}$  & $6.83$ & $3.20\times 10^{-5}$ & $1.14\times 10^{-2}$ & $6.97$& $1.00\times 10^{-7}$ & $9.71\times 10^{-7}$ \\
\hline
F10\_c  & $2.25\times 10^{9}$  & $12.01$ & $3.42\times 10^{-4}$ & $3.78\times 10^{-2}$ & $12.20$& $1.00\times 10^{-7}$ & $6.60\times 10^{-6}$ \\
\hline
F11\_a  & $3.75\times 10^{9}$  & $10.37$ & $1.04\times 10^{-4}$ & $1.78\times 10^{-3}$ & $10.43$ & $1.00\times 10^{-6}$ & $1.04\times 10^{-5}$ \\
\hline
F11\_b  & $3.75\times 10^{9}$  & $15.90$ & $1.87\times 10^{-3}$ & $5.26\times 10^{-3}$ & $15.96$ & $1.00\times 10^{-7}$ & $1.80\times 10^{-5}$ \\
\hline
F12\_a  & $1.87\times 10^{9}$  & $6.18$ & $1.00\times 10^{-7}$ & -- & $6.34$& $1.00\times 10^{-7}$ &-- \\
\hline
F12\_b  & $1.87\times 10^{9}$  & $10.04$ & $3.00\times 10^{-4}$ & -- & $10.21$& $1.00\times 10^{-7}$ &-- \\
\hline
F12\_c  & $1.87\times 10^{9}$  & $16.96$ & $5.30\times 10^{-3}$ & -- & $17.15$& $4.00\times 10^{-6}$
&-- \\
\hline

F13\_a  & $2.62\times 10^{10}$  & $2517.40$ & $6.36\times 10^{1}$ & $5.61$ & $2515.49$ & $5.49\times 10^{-4}$ & $2.56\times 10^{-3}$ \\
\hline
F13\_b  & $2.63\times 10^{11}$  & $108239.15$ & $1.46\times 10^{2}$ & $ 8.50 $ & $108113.13$ & $9.52\times 10^{-3}$ & $1.00\times 10^{-2}$ \\
\hline
F14  & $1.25\times 10^{9}$  & $53.53$ & $5.37$ & $4.63$ & $53.06 $& $1.00\times 10^{-6}$ & $1.11\times 10^{-3}$
\\
\hline
F15  & $3.39\times 10^{13}$  & $52220.44$ & $4.96\times 10^{-3}$ & $1.35\times 10^{-2}$ & $52008.19$ & $1.00\times 10^{-7}$ & $1.45\times 10^{-6}$ \\
\hline
F16  & $1.87\times 10^{9}$  & $4.03$ & $4.53\times 10^{-7}$ & $4.35\times 10^{-4}$ & $4.18$ & $4.53\times 10^{-7}$ & $6.53\times 10^{-5}$ \\
\hline
F17  & $1.87\times 10^{9}$  & $5.87$ & $1.00\times 10^{-7}$ & $1.20\times 10^{-5}$ & $5.92$ & $1.00\times 10^{-7}$ & $2.20\times 10^{-6}$\\
\hline
F18\_a  & $1.87\times 10^{9}$  & $7.36$ & $1.49\times 10^{-4}$ & $1.36\times 10^{-4}$ & $7.46$ & $1.00\times 10^{-7}$ & $2.00\times 10^{-5}$ \\
\hline
F18\_b  & $1.87\times 10^{9}$  & $8.89$ & $1.25\times 10^{-4}$ & $1.84\times 10^{-4}$ & $9.00$ & $1.00\times 10^{-7}$ &  $1.39\times 10^{-4}$ \\
\hline
F18\_c  & $1.87\times 10^{9}$  & $11.19$ & $3.11\times 10^{-4}$ & $3.30\times 10^{-4}$ & $11.31$ & $1.00\times 10^{-7}$ & $1.47\times 10^{-4}$ \\
\hline
F19\_a  & $1.87\times 10^{9}$  & $17.09$ & $1.00\times 10^{-7}$ & $1.07\times 10^{-5}$ & $17.58$ & $4.00\times 10^{-6}$ & $4.92\times 10^{-6}$ \\
\hline
F19\_b  & $1.87\times 10^{9}$  & $33.52$ & $2.10\times 10^{-3}$ & $2.89\times 10^{-4}$ & $33.96$ & $4.00\times 10^{-6}$ & $4.61\times 10^{-6}$ \\
\hline
\end{tabular}
}
\caption{\footnotesize Results for the test problem suite. In the column Error we indicate the relative error in $ || \cdot ||_2$ when the location of the minimum is non zero, otherwise the absolute error is presented. Cells with '-' mark correspond to cases in which the exact minima are unknown. }
\label{academico2}
\end{table}

\begin{table}[!htb]
\centering
{\tiny
\begin{tabular}{|r||l|r|l|l||r|l|l|}
\hline
$f$    & \multicolumn{4}{c||}{\bf V2} & \multicolumn{3}{c|}{\bf Hybrid}\\
\hline
        & Function eval.  & Time & $|f_{a}-f_{r}|$ & Error & Time & $|f_{a}-f_{r}|$ & Error \\
\hline
F0\_g  & $5.40 \times 10^{7}$  & $31.13 $ & $5.10 $ & $ 1.51\times 10^{-2}$ & $2.24$ & $ 2.10\times 10^{-12} $ & $1.01\times 10^{-8} $ \\
\hline
F1\_d  & $8.33\times 10^{7}$  & $ 36.96$ & $1.53 $ & $3.67 $ & $0.79$ & $ 2.17\times 10^{-8}$ & $ 1.50\times 10^{-12} $ \\
\hline
F8\_c  & $ 9.01\times 10^{7}$  & $ 81.26$ & $1.38\times 10^{-1} $ & $ 6.91$  & $1.44 $ & $3.33\times 10^{-16} $ & $1.08\times 10^{-6}$ \\
\hline
F13\_b & $3.47\times 10^{8}$  & $165.57 $ & $2.36\times 10^{1} $ & $3.45\times 10^{-1} $  & $1.40 $ & $3.63\times 10^{-12} $ & $ 2.44\times 10^{-7}$ \\
\hline
\end{tabular}
}
\caption{\footnotesize Results of the hybrid algorithm. The first part shows the results of the annealing algorithm. The second one shows the results of the Nelder-Mead algorithm starting at the point at which the annealing algorithm was stopped prematurely.}
\label{Tabla:hybrid}
\end{table}

\section{Conclusions}

The extremely long execution times associated to SA algorithm in its sequential version results to be its main drawback when applied to realistic optimization problems that involve high dimension spaces or function evaluations with high computational cost. This is the reason why many authors in the literature have designed different alternatives to parallelize sequential SA by using different high-performance computing techniques. In the present paper we have developed an efficient implementation of a SA algorithm by taking advantage of the power of GPUs. After analyzing a sequential SA version, a straightforward asynchronous and a synchronous implementations have been developed, the last one including an appropriate communication among Markov chains at each temperature level. The parallelization of the SA algorithm in GPUs has been discussed and the convergence of the different parallel techniques has been analyzed. Moreover, the parallel SA algorithm implementations have been checked by using classical experiments. A deeper analysis of results for a model example problem is detailed and the list of test examples defining the benchmark is included in Appendix. In summary, the results illustrate a better performance of the synchronous version in terms of convergence, accuracy and computational cost. Moreover, some results illustrate the behavior of the SA algorithm when combined with the Nedler-Mead local minimization method as an example of hybrid strategy that adequately balances accuracy and computational cost in real applications. The resulting code is planned to be leveraged in open source. \\

\appendix
\section{Appendix: Test functions}

In this appendix we present the expressions of the functions in our test problem suite.

\begin{enumerate}

\item {\bf Test 1} ({\it Ackley problem}):

Originally the \emph{Ackley's} problem \cite{ACK87} was defined for two
dimensions, but the problem has been generalized to $n$ dimensions \cite{BAC96}.

Formally, this problem can be described as finding a point
$\pmb x=(x_{1},x_{2},\ldots{},x_{n})$, with $x_{i}\in [-30,
30]$, that minimizes the following equation:

\begin{equation*}
    \mathop{f(\pmb x)}=-20 \exp\left(-0.2\sqrt{\frac{1}{n} \sum\limits_{i=1}^{n}x_{i}^{2}}\,\right)
                    - \exp\left(\frac{1}{n} \sum\limits_{i=1}^{n}\cos(2\pi
                    x_{i})\right) + 20 + e.
\end{equation*}

\noindent The minimum of the Ackley's function is located at the origin with
$f(\pmb 0)=0$. This test was performed for $n=30$, $n=100$, $n=200$
and $n=400$.

\item {\bf Test 2} ({\it Branin problem}):

The expression of the Branin function (\cite{Dixon-Szego-78}) is,

$$
f(\pmb x)= \left(x_2- \frac{5.1}{4 \pi^2}x_1^2+ \frac{5}{\pi} x_1-6\right) ^2
+
10\left(1-\frac{1}{8\pi} \right)\cos(x_1) + 10,
$$

\noindent with $x_1, \, x_2 \in [-20,20]$. The minimum of the objective function value is equal to
$f(\pmb x^{\star})=0.397887$, and its located at the following three points: $\pmb
x^{\star}=(-\pi, 12.275)$, $\pmb
x^{\star}=(\pi, 2.275)$ and $\pmb
x^{\star}=(9.425, 2.475)$.

% funci?n 17 -- pag 288 -- articulo 2007

\item {\bf Test 3} ({\it Cosine mixture problem}):

The expression of this function is \cite{BreimanCutler-93}:

$$
f(\pmb x)= -\displaystyle 0.1 \sum_{i=1}^n \cos(5 \pi x_i) - \sum_{i=1}^n x_i^2,
$$

\noindent with $x_i \in [-1,1]$, $i = 1,\, 2 \, \ldots, n$. The global minimum
is located at the origin with the function values $-0.2$ and $-0.4$ for $n=2$ and
$n=4$, respectively.

\item {\bf Test 4} ({\it Dekkers and Aarts problem}):

The Dekkers and Aarts function (\cite{DekkersAarts-91}) has the following
expression

$$
f(\pmb x)= 10^5 x_1^2+x_2^2-(x_1^2+ x_2^2)^2+ 10^{-5}(x_1^2+x_2^2)^4,
$$

\noindent with $x_1, \, x_2 \in [-20,20]$. This function has more than three
local minima, but there are two global minima located at $\pmb
x^{\star}=(0,-14.945)$ and $\pmb x^{\star}=(0,14.945)$ with $f(\pmb x^{\star})=-24776.518$.

\item {\bf Test 5} ({\it Easom problem}):

The Easom function (\cite{Michalewicz98}) has the following definition

$$
f(\pmb x)=-\cos(x_1) \cos(x_2)\exp( - (x_1- \pi) ^2 - (x_2- \pi)^2),
$$
\noindent where the considered search space is  $x_1, \, x_2  \in [-10,10]$. The
minimum value is located at $\pmb x^{\star}=(\pi, \pi)$ with $f(\pmb
x^{\star})=-1$.

\item {\bf Test 6} ({\it Exponential problem}):

The definition of the Exponential problem (see \cite{BreimanCutler-93}) is the
following

$$
f(\pmb x)=- \exp\left( -0.5 \displaystyle \sum_{i=1}^n x_i^2 \right),
$$

\noindent with $x_i\in [-1,1]$, $i=1,\ldots, n$. The optimal objective function value is
$f(\pmb x^{\star})=-1$, and it's located at the origin. In our tests we consider $n=4$.

\item {\bf Test 7} ({\it Goldstein and Price problem}):

The Goldstein and Price function (see \cite{Dixon-Szego-78}) has the following
definition,

$$
\begin{array}{rl}
f(x)= & [1+(x_1+x_2+1)^2(19-14x_1+3x_1^2-14x_2+6x_1x_2)+3x_2^2]\\
\, &\times[30+(2x_1-3x_2)^2(18-32x_1+12x_1^2+48x_2-36 x_1 x_2+ 27 x_2^2)],
\end{array}
$$

\noindent with $x_1, \, x_2 \in [-2,2]$. There are four local minima and the
global minimum is located at $\pmb x^{\star}= (0,-1)$ with $f(\pmb x^{\star})=3$.

\item {\bf Test 8} ({\it Griewank problem}):

The Griewank function (proposed in \cite{Griewank81}) is defined as
follows,

$$
f(\pmb x) = 1+ \displaystyle \sum_{i=1}^n \left[ \frac{x_i^2}{4000}- \prod_{i=1}^{n}
\cos\left( \frac{x_i}{\sqrt{i}}\right)\right] ,
$$

\noindent where $x_i \in [-600, 600]$, $i=1, \ldots, n$. The global minimum
is located at the origin and it's function value is $0$; moreover the function also has a very
large number of local minima, exponentially increasing with $n$ (in the two
dimensional case there are around $500$ local minima). Tests were performed for
$n=100$, $n=200$ and $n=400$.

\item {\bf  Test 9} ({\it Himmelblau problem}):

The expression of the Himmelblau's function (\cite{HimmelBlau-72}) is the
following

$$
f(\pmb x)= (x_1^2 + x_2 -11)^2+ (x_1+x_2^2-7)^2,
$$

\noindent where $x_1,  \, x_2 \in [-6,6]$. The global minima is located at the following four points $\pmb x^{\star}=(3.0,2.0)$, $\pmb x^{\star}=(-2.805118, 3.131312)$, $\pmb x^{\star}=(-3.779310,-3.283186)$ and
$\pmb x^{\star}=(3.584428,-1.848126)$, with $f(\pmb x^{\star}) = 0$.

\item {\bf  Test 10} ({\it Levy and Montalvo problem}):

The expression of the Levy and Montalvo function (\cite{LevyMontalvo-1985}) is,

$$
f(\pmb x)= \frac{\pi}{n} \left(
10 \sin^2(\pi y_1) + \displaystyle \sum_{i=1}^{n-1}(y_i-1)^2 \left(1+10 \sin^2(\pi
y_{i+1})\right) + (y_n-1)^2
\right),
$$

\noindent where $y_i= 1+ \frac{1}{4} (x_i+1)$ for $x_i \in [-10,10]$, $i=1,
\ldots, n$. This function has approximately $5^n$ local minima and the global
minimum is  located at the point $\pmb x^{\star}= (-1,
\ldots, -1)$ with $f(\pmb x^{\star})=0$. Tests were performed for $n=2$, $n=5$ and $n=10$.

 \item {\bf  Test 11} ({\it Modified Langerman problem}):

The expression of the Modified Langerman function \cite{LangermanFunct-1996} is,

\begin{equation*}
f(\pmb x) = - \sum_{i=1}^{5} c_i \left[ \exp \left(-\frac{1}{\pi} \sum_{j=1}^{n} (x_j-a_{ij})^2 \right)
\cos \left( \pi \sum_{j=1}^{n}(x_j-a_{ij})^2 \right) \right],
\end{equation*}
\noindent where $x_i\in [0,10]$, $i=1, \ldots, n$ and
\begin{displaymath}
\pmb A = (a_{ij}) =
\left(
\begin{array}{cccccccccc}
9.681 & 0.667 & 4.783 & 9.095 & 3.517 & 9.325 & 6.544 & 0.211 & 5.122 & 2.020 \\
9.400 & 2.041 & 3.788 & 7.931 & 2.882 & 2.672 & 3.568 & 1.284 & 7.033 & 7.374 \\
8.025 & 9.152 & 5.114 & 7.621 & 4.564 & 4.711 & 2.996 & 6.126 & 0.734 & 4.982 \\
2.196 & 0.415 & 5.649 & 6.979 & 9.510 & 9.166 & 6.304 & 6.054 & 9.377 & 1.426 \\
8.074 & 8.777 & 3.467 & 1.863 & 6.708 & 6.349 & 4.534 & 0.276 & 7.633 & 1.567 \\
\end{array}
\right),
\end{displaymath}
\begin{displaymath}
\pmb c = (c_i)=
\left(
\begin{array}{ccccc}
0.806 & 0.517 & 0.100 & 0.908 & 0.965 \\
\end{array}
\right).
\end{displaymath}

\noindent  In this case it is unknown
the number of local minima. The global optimum with $n=2$ is searched at
$\pmb x^{\star}=(9.6810707,0.6666515)$ with $ f(\pmb x^{\star})=-1.080938$, and for $n=5$ the
global minimun is located at $\pmb x^{\star}=(8.074000, 8.777001,3.467004,
1.863013, 6.707995)$ with $ f(\pmb x^{\star})=-0.964999$.

\item {\bf  Test 12} ({\it Michalewicz problem}):

The definition of the Michalewicz function (\cite{Michalewicz98}) is the
following,

$$
f(\pmb x)= - \displaystyle \sum_{i=1}^n \sin (x_i) \left[
\sin \left(
\frac{i x_i^2}{\pi}
\right)
\right]^{2m}, \quad
$$

\noindent where $x_i \in [0, \pi]$, $i= 1,\ldots, n$. It is usually set $m=10$.
The objective function value at the global minimum is $f(\pmb x^{\star})=-1.8013$ for $n=2$, $f(\pmb
x^{\star})=-4.6877$ for $n=5$, and $f(\pmb x^{\star})=-9.6602$ for $n=10$.

\item {\bf  Test 13} ({\it Rastrigin problem}):

The expression of the Rastrigin function (see \cite{StornPrice-97} and
\cite{TormZilinska-1996}, for example) has the following definition,

$$
f(\pmb x)= 10 n + \displaystyle \sum_{i=1}^{n} \left(
x_i^2 - 10 \cos (2 \pi x_i)
\right),
$$

\noindent where $x_i \in [-5.12, 5.12]$, $i=1, \ldots, n$. The global minimum
is located at $\pmb x^{\star}= (0,\ldots, 0)$ and the objective function at this point is
 $f(\pmb x^{\star})=0$. In our tests we consider $n=100$ and $n=400$.

\item {\bf  Test 14} ({\it Generalized Rosenbrock problem}):

The Rosenbrock's function (\cite{DeJohnThesis-75}), also known as
Rosenbrock valley, banana function or the {\it second function of De Jong}, has
the following expression,

$$
f(\pmb x)= \displaystyle \sum_{i=1}^{n-1} \left[
100 (x_{i+1}- x_i)^2 + (1-x_i)^2
\right],
$$

\noindent with $x_i \in [-2.048, 2.048]$, $i=1, \ldots, n$. The global minimum
 is located at $\pmb x^{\star}=(1, \ldots,1)$ with the function value $f(\pmb x^{\star})=0$.  In our tests we consider $n=4$.

\item {\bf  Test 15} ({\it Salomon problem}):

The Salomon function (\cite{Salomon95}) has the following definition,

$$
f(\pmb x)= 1 - \cos (2 \pi || \pmb x ||_2 )+ 0.1 || \pmb x ||_2,
$$

\noindent where $|| \pmb x ||_2=\sqrt{\displaystyle \sum_{i=1}^n x_i^2}$, and
$x_i \in [-100, 100]$, $i=1, \ldots, n$. In the general case ($n$) is unknown
the number of local minima. It has a global minima located at $\pmb
x^{\star}=(0,\ldots, 0)$ with $f(\pmb x^{\star})=0$. For our tests we consider
$n=10$.

\item {\bf  Test 16} ({\it Six-Hump Camel Back problem}):

The expression of the Six-Hump Camel Back function (\cite{Dixon-Szego-78}) is
the following,

$$
f(\pmb x)= \left(
4-2.1x_1^2+ \frac{1}{3}x_1^4
\right)x_1^2+ x_1 x_2 + (-4 + 4 x_2^2)x_2^2,
$$

\noindent with $x_1 \in [-3,3]$ and $x_2 \in [-2,2]$. This function has two
global minima equal to $f(\pmb x^{\star})=-1.0316$, located at $\pmb
x^{\star}=(-0.0898,0.7126)$ and $\pmb
x^{\star}=(0.0898, -0.7126)$.

\item {\bf  Test 17} ({\it Shubert problem}):

The Shubert function (\cite{LevyMontalvo-1985}) has the
following definition

$$
f(\pmb x)= \displaystyle \prod_{i=1}^n \left(
\sum_{j=1}^{5} j \cos( (j+1)x_i +j)
\right),
$$

\noindent subject $x_i \in [-10,10]$, $i=1,\ldots,n$. For the $n$-dimensional
case the number of local minima is unknown, however for $n=2$, the function has
$760$ local minima, where $18$ of them are global with $f(\pmb
x^{\star})\approx -186.7309$. We have performed the tests for $n=2$. For this case, the global optimizers are $(-7.0835,4.8580), (-7.0835,-7.7083), (-1.4251,-7.0835), (5.4828,4.8580), (-1.4251,-0.8003)$ $(4.8580,5.4828), (-7.7083,-7.0835), (-7.0835,-1.4251), (-7.7083,-0.8003), (-7.7083,5.4828), $ $(-0.8003,-7.7083), (-0.8003,-1.4251), (-0.8003,4.8580), (-1.4251,5.4828), (5.4828,-7.7083), $ $ (4.8580,-7.0835), (5.4828,-1.4251), (4.850,-0.8003)$.

\item {\bf  Test 18} ({\it Shekel problem}):

The expression of the Shekel function (\cite{LangermanFunct-1996}) is
\begin{equation*}
 \displaystyle f(\pmb x)=- \displaystyle \sum_{i=1}^{m} \dfrac{1}{
\displaystyle \sum_{j=1}^{4} (x_j-a_{ij})^2+c_i},
\end{equation*}
where the matrix $A=(a_{ij})$ and the vector $\pmb c= (c_i)$ are the following,
{\footnotesize
$$
\begin{array}{cc}
\pmb c= \left(
\begin{array}{c}
0.1\\
0.2\\
0.2\\
0.4\\
0.4\\
0.6\\
0.3\\
0.7\\
0.5\\
0.5\\
\end{array}
\right),
&
A=
\left(
\begin{array}{ccccc}
4 & 4 & 4 & 4 \\
1 & 1 & 1 & 1 \\
8 & 8 & 8 & 8 \\
6 & 6 & 6 & 6 \\
3 & 7 & 3 & 7 \\
2 & 9 & 2 & 9 \\
5 & 5 & 3 & 3 \\
8 & 1 & 8 & 1 \\
6 & 2 & 6 & 2 \\
7 & 3.6 & 7 & 3.6 \\
\end{array}
\right).
\end{array}
$$
}

The search domain is $x_i \in [0,10]$, $i=1,\ldots,4$. The global optimum is $\pmb x^{\star}=(4, 4, 4, 4)$ and the function value at this point is $f(\pmb x^{\star})= -10.1532$ for $m=5$, $f(\pmb x^{\star})= -10.4029$ for $m=7$ and $f(\pmb x^{\star})= -10.5364$ for $m=10$.

\item {\bf  Test 19} ({\it Modified Shekel Foxholes problem}):

The expression of the Modified Shekel Foxholes function (\cite{LangermanFunct-1996}) is

\begin{equation*}
 \displaystyle f(\pmb x)=- \displaystyle \sum_{i=1}^{30} \dfrac{1}{
\displaystyle \sum_{j=1}^{n} (x_j-a_{ij})^2+c_i},
\end{equation*}
where the matrix $A=(a_{ij})$ and the vector $\pmb c= (c_i)$ are the following,
{\footnotesize
$$
\begin{array}{cc}
\pmb c= \left(
\begin{array}{c}
0.806\\
0.517\\
0.100\\ 0.908\\
0.965\\  0.669\\
0.524\\  0.902\\
0.531\\  0.876\\
0.462\\  0.491\\
0.463\\  0.714\\
0.352\\ 0.869\\
0.813\\  0.811\\
0.828\\
0.964\\
0.789\\
0.360\\
0.369\\
0.992\\
0.332\\
0.817\\
0.632\\
0.883\\
0.608\\
0.326\\
\end{array}
\right),
&
A=
\left(
\begin{array}{cccccccccc}
9.681 & 0.667 & 4.783 & 9.095 & 3.517 & 9.325 & 6.544 & 0.211 & 5.122 & 2.020\\
9.400  & 2.041  & 3.788 &  7.931 &  2.882 &  2.672 & 3.568 & 1.284 & 7.033 &
7.374\\
8.025  & 9.152  & 5.114  & 7.621  & 4.564  & 4.711 &  2.996  & 6.126  & 0.734
 & 4.982\\
2.196  & 0.415  & 5.649  & 6.979  & 9.510  & 9.166 &  6.304 &  6.054  & 9.377
 & 1.426\\
8.074  &  8.777  & 3.467  & 1.863  & 6.708  & 6.349  & 4.534  & 0.276 &  7.633
 & 1.567\\
7.650  & 5.658  & 0.720 &  2.764  & 3.278  & 5.283  & 7.474  & 6.274  & 1.409 &
8.208\\
1.256  & 3.605  & 8.623 &  6.905  & 4.584  & 8.133  & 6.071  & 6.888  & 4.187 &
5.448\\
8.314  & 2.261 &  4.224  & 1.781  & 4.124 &  0.932  & 8.129  & 8.658  & 1.208
 & 5.762\\
0.226  & 8.858  & 1.420  & 0.945  & 1.622 &  4.698 &  6.228  & 9.096 &  0.972 &
7.637\\
7.305  & 2.228  & 1.242  & 5.928  & 9.133  & 1.826  & 4.060  & 5.204  & 8.713 &
8.247\\
0.652 &  7.027 &  0.508 &  4.876  & 8.807  & 4.632  & 5.808 &  6.937 &  3.291
 & 7.016\\
2.699 &  3.516  & 5.874  & 4.119  & 4.461  & 7.496 &  8.817  & 0.690  & 6.593
 & 9.789\\
8.327  & 3.897 &  2.017  & 9.570 &  9.825 &  1.150 &  1.395 &  3.885 &  6.354
 & 0.109\\
2.132  & 7.006  & 7.136  & 2.641 &  1.882  & 5.943 &  7.273 &  7.691 &  2.880
 & 0.564\\
4.707  & 5.579 &  4.080 &  0.581 &  9.698 &  8.542 &  8.077 &  8.515 &  9.231
 & 4.670\\
8.304  & 7.559  & 8.567  & 0.322  & 7.128  & 8.392  & 1.472  & 8.524  & 2.277
 & 7.826\\
8.632 &  4.409 &  4.832 &  5.768  & 7.050 &  6.715 &  1.711 &  4.323 &  4.405
 & 4.591\\
4.887  & 9.112  & 0.170 &  8.967 &  9.693  & 9.867 &  7.508 &  7.770 &  8.382
 & 6.740\\
2.440  & 6.686  & 4.299 &  1.007 &  7.008  & 1.427  & 9.398  & 8.480  & 9.950
 & 1.675\\
6.306  & 8.583 &  6.084 &  1.138 &  4.350 &  3.134 &  7.853 &  6.061 &  7.457
 & 2.258\\
0.652 &  2.343  & 1.370 &  0.821 &  1.310  & 1.063 &  0.689  & 8.819 &  8.833 &
9.070\\
5.558  & 1.272 &  5.756 &  9.857  & 2.279  & 2.764 &  1.284  & 1.677  & 1.244
 & 1.234\\
3.352  & 7.549  & 9.817  & 9.437 &  8.687  & 4.167  & 2.570 &  6.540  & 0.228
 & 0.027\\
8.798  & 0.880 &  2.370 &  0.168 &  1.701 &  3.680  & 1.231  & 2.390 &  2.499
 & 0.064\\
1.460  & 8.057  & 1.336 &  7.217  & 7.914  & 3.615 &  9.981 &  9.198 &  5.292
 & 1.224\\
0.432  & 8.645  & 8.774  & 0.249  & 8.081  & 7.461  & 4.416  & 0.652  & 4.002
 & 4.644\\
0.679  & 2.800 &  5.523  & 3.049 &  2.968  & 7.225 &  6.730 &  4.199 &  9.614
 & 9.229\\
4.263  & 1.074  & 7.286 &  5.599 &  8.291  & 5.200  & 9.214  & 8.272  & 4.398
 & 4.506\\
9.496  & 4.830  & 3.150 &  8.270  & 5.079 &  1.231 &  5.731  & 9.494 &  1.883
 & 9.732\\
4.138  & 2.562  & 2.532 &  9.661  & 5.611 &  5.500  & 6.886  & 2.341  & 9.699
 & 6.500
\end{array}
\right).
\end{array}
$$
}

\noindent The search domain is $x_i \in [-5,15]$. For this function the number of local minima is unknown.
For $n=2$ the global minimum is located at the
point $\pmb x^{\star}=(8.024, 9.146)$ with $f(\pmb x^{\star})= -12.1190$. For $n=5$ the global minima is $\pmb x^{\star}=(8.025, 9.152, 5.114, 7.621, 4.564)$ with $f(\pmb x^{\star})= -10.4056$.

\end{enumerate}

\end{document}